\documentclass[a4paper,11pt]{article}
\pdfoutput=1 

\usepackage{jcappub} 

\usepackage[T1]{fontenc} 
\usepackage[varg]{txfonts}
\usepackage{bm}
\usepackage{textcomp}
\usepackage{gensymb}
\usepackage[dvipsnames]{xcolor}
\usepackage{hyperref}
\usepackage[printwatermark]{xwatermark}
\def\be{\begin{equation}}
\def\ee{\end{equation}}
\def\ba{\begin{eqnarray}}
\def\ea{\end{eqnarray}}
\def\eqi{\begin{equation}}
\def\eqf{\end{equation}}
\def\eqia{\begin{eqnarray}}
\def\eqfa{\end{eqnarray}}
\def\lcdm{$\Lambda$CDM }
\def\nn{\nonumber}

\title{Constraints on the distance duality relation with standard sirens}

\author[a]{N. B. Hogg,}
\author[b]{M. Martinelli,}
\author[b]{and S. Nesseris}

\affiliation[a]{Institute  of  Cosmology  and  Gravitation,  University  of  Portsmouth,\\ Burnaby  Road,  Portsmouth  PO1  3FX, United Kingdom}
\affiliation[b]{Instituto de F\'isica T\'eorica UAM-CSIC,\\ Campus de Cantoblanco, 28049 Madrid, Spain}

\emailAdd{natalie.hogg@port.ac.uk}
\emailAdd{matteo.martinelli@uam.es}
\emailAdd{savvas.nesseris@csic.es}

\abstract{We use gravitational wave (GW) standard sirens, in addition to Type Ia supernovae (SNIa) and baryon acoustic oscillation (BAO) mock data, to forecast constraints on the electromagnetic and gravitational distance duality relations (DDR). We make use of a parameterised approach based on a specific DDR violation model, along with a machine learning reconstruction method based on the Genetic Algorithms. We find that GW provide an alternative to the use of BAO data to constrain violations of the DDR, reaching $3\%$ constraints on the violation parameter we consider when combined with SNIa, which is only improved by a factor of $\approx1.4$ if one instead considers the combination of BAO and SNIa. We also investigate the possibility that a neglected modification of gravity might lead to a false detection of DDR violations, even when screening mechanisms are active. We find that such a false detection can be extremely significant, up to $\approx10\sigma$ for very extreme modified gravity scenarios, while this reduces to $\approx4\sigma$ in a more realistic case. False detections can also provide a smoking gun for the modified gravity mechanism at play, as a result of the tension introduced between the SNIa+GW and SNIa+BAO combinations.}


\begin{document}
\maketitle
\flushbottom

\section{Introduction}
\label{sec:intro}

On the $11^{\rm th}$ February 2016, the Laser Interferometer Gravitational-Wave Observatory (LIGO) and Virgo collaborations announced the first direct detection of a gravitational wave signal, coming from the inspiral, merger and ringdown of a binary black hole system \cite{GWthefirst}. The subsequent observation in 2017 of a binary neutron star merger concurrent with an electromagnetic counterpart (GW170817) heralded a new era of multi-messenger astronomy and the use of gravitational wave events as so-called ``standard sirens'' \cite{Schutz1986, Holz:2005df, Dalal:2006qt,BNSmerger}.

These two groundbreaking observations had important repercussions for cosmology too. The first lent further support to Einstein's theory of general relativity (GR) by confirming the theory's prediction of gravitational waves; the second placed strong constraints on modified gravity theories that predicted a tensor speed different to that of light \cite{Monitor:2017mdv,Ezquiaga:2017ekz,Creminelli:2017sry}, as well as providing a new distance ladder independent measurement of the Hubble parameter $H_0$ \cite{Abbott:2017xzu}. Such a measurement of $H_0$ is still not competitive with those provided by other probes \cite{Chen:2017rfc}, but it highlights how future GW telescopes will be able to shed light on the cosmological tension problems faced by contemporary cosmology. Recent investigations have also shown how the observation of GW will provide new tests of General Relativity, potentially constraining several still viable modified gravity theories (see e.g. \cite{Mukherjee:2019wcg,Mukherjee:2019wfw,Mukherjee:2020hyn}).

One example of a future gravitational wave observatory is the Einstein Telescope (ET), a proposed ground-based triangular interferometer that will be part of the third generation of gravitational wave detectors \cite{Maggiore:2019uih}. 
Current terrestrial detectors such as LIGO and Virgo are limited in the low frequency range by seismic and thermal noise; these factors will be mitigated especially in the case of the ET by its proposed underground construction and cryogenic cooling of the interferometer mirrors. The reduced noise levels of the ET and other third generation detectors will therefore enable extremely sensitive measurements of gravitational wave signals to be made, bringing standard siren detections into the realm of precision cosmology \cite{Hild:2010id}. 

As our measurements of cosmological parameters improve, the standard cosmological model of a spatially flat Universe dominated by a cosmological constant plus cold dark matter ($\Lambda$CDM) is still the most appealing to explain observations with respect to the most common alternatives (see e.g. \cite{Aghanim:2018eyx,Abbott:2018xao} for recent constraints obtained by the Planck and DES surveys). Therefore, it becomes necessary to consider how best to constrain more exotic deviations from the standard paradigms of general relativity and $\Lambda$CDM. A feasible possibility is a violation of the distance duality relation (DDR), which relates angular diameter and luminosity distances, a possibility for which several observational tests have been proposed \cite{Holanda:2010vb,Avgoustidis:2010ju,Holanda:2012at,Liao:2015ccl,Liao:2015uzb,Liao:2019xug}. Deviations can occur in both the electromagnetic (EM) and gravitational wave (GW) sectors. However, these would be due to very different physical mechanisms, with the former related to a non-standard propagation of photons and the latter to an anomalous propagation of gravitational waves.

In this work, we focus on the first of these possibilities, studying a toy model in which the EM DDR is broken as photons decay into axions while propagating through cosmic magnetic fields. Such deviations of the DDR are commonly constrained using SNIa observations alongside BAO measurements, where the latter are not sensitive to the violation mechanisms and can therefore break the degeneracies between DDR violation and standard cosmological parameters. We explore the possible use of future GW datasets as an alternative to BAO, or alongside them, to constrain the DDR violation model under examination. We exemplify our method using mock datasets for future observations of SNIa, BAO and GW and, using an MCMC analysis, show the constraints that can be obtained on cosmological and model parameters. However, this approach can lead to false detections of DDR violations if mechanisms leading to anomalous GW propagation are also at play and are not considered in the analysis. Exploring this possibility, we attempt to highlight the signatures of such a scenario in the final results of the analysis pipeline, investigating the constraints one would obtain if both deviations from the standard behaviour are considered at the same time.

Finally, we also perform a machine learning reconstruction of the distance duality relations as functions of redshift, using Genetic Algorithms (GA). The GA are a stochastic optimisation approach that, given some data, can provide functional reconstructions that depend solely on the redshift $z$ and are based on a minimal set of assumptions \cite{Bogdanos:2009ib,Nesseris:2012tt}. The main advantage of this approach is that the GAs are not susceptible to theoretical priors about the behaviour of the data under question and can detect hidden features in the data, that at first sight might be missed by traditional inference approaches. The GA can also help avoid biases in the results and possible false detections of DDR violations, something which we explicitly test using mock data in order to validate our approach.

This paper is organised as follows: in Section \ref{sec:theory} we describe deviations from the standard DDR in the electromagnetic sector focusing on a toy model describing the decay of photons into axions, and in Section \ref{sec:LCDM} we explicitly demonstrate how standard sirens can be used as an alternative to, or alongside, BAO to constrain these deviations, briefly describing the mock datasets we use for this analysis; in Section \ref{sec:MG} we extend our analysis of DDR violations when anomalous GW propagation is also present, here due to a modified gravity mechanism, and investigate the bias that such a mechanism would introduce in our results; in Section \ref{sec:GA} we show the results of our GA reconstruction of the DDR; in Section \ref{sec:conclusion} we present our conclusions. Finally, in Appendix \ref{sec:mockdata} we provide more technical details related to our mock datasets, describing the specifications of the various surveys and telescopes we consider and the procedure we followed to create them.

\section{Photon decay and deviations from standard DDR}\label{sec:theory}


The investigation of the homogeneous expansion of the Universe commonly relies on the observations of standard candles, which probe the luminosity distance $d_L$, and standard rulers, through which we can measure the angular diameter distance $d_A$. The general relation between these quantities, which holds under the two conditions that the number of photons is conserved and that they travel on null geodesics in a pseudo-Riemannian spacetime \cite{Ellis2007, Bassett:2003vu}, is given by
\begin{equation}\label{eq:std_DDR}
    d_L(z)=(1+z)^2d_A(z),
\end{equation}
which is called the distance duality relation (DDR). Both the luminosity and angular diameter distance can be obtained in terms of the comoving distance $r(z)$ as
\begin{align}
 d_L(z)=&(1+z)r(z)\, , \label{eq:dl}\\
 d_A(z)=&\frac{r(z)}{1+z}\, .\label{eq:da}
\end{align}
Even though these relations hold for the minimal set of assumptions mentioned above, in this paper we assume that the background expansion of the Universe is the one produced by a spatially flat $\Lambda$CDM model, which allows the comoving distance to be expressed as
\begin{equation}\label{eq:chi}
    r(z) = c\int_0^z{\frac{dz'}{H(z')}}\, ,
\end{equation}
where $H(z)$ is the Hubble parameter in units of km s$^{-1}$ Mpc$^{-1}$ and $c$ is the speed of light in km s$^{-1}$.

In this paper we focus on a violation of the first condition, photon number conservation, investigating mechanisms that lead photons to be converted into other particles, such as axions or other axion-like particles \cite{Tiwari:2016cps}. Axion models have received a spike in interest after the recent XENON1T observation of excess electronic recoil, which was attributed to solar axions with a significance of $3.5 \sigma$ \cite{Aprile:2020tmw}\footnote{It has been noted that astrophysical constraints on solar axions are incompatible with the XENON1T excess \cite{DiLuzio:2020jjp}, and that the detection could be due to the previously unaccounted-for $\beta$ decays of tritium in the detector \cite{Robinson:2020gfu}. The significance of the solar axion fit decreases to $2.1\sigma$ if this additional tritium component is considered.}. Here we examine a specific mechanism that considers the possibility of novel scalar and pseudo-scalar particles inspired from beyond Standard Model (BSM) physics coupling to the photons via the following interaction terms \cite{Avgoustidis:2010ju}
\be
\mathcal{L}_{\rm int,scalar}=\frac{1}{4M}F_{\mu \nu}F^{\mu \nu}\phi \label{eq:axion_coupling_1}
\ee
and
\be
\mathcal{L}_{\rm int,pseudo}=\frac{1}{8M}\epsilon_{\mu \nu \lambda\rho}F^{\mu \nu}F^{\lambda\rho}\phi, \label{eq:axion_coupling_2}
\ee
where $M$ is the energy scale of the coupling, $\epsilon_{\mu \nu \lambda\rho}$ the Levi-Civita antisymmetric symbol, $\phi$ is the axion particle and $F^{\mu \nu}$ the electromagnetic field strength. In the presence of  magnetic fields, photons have a non-vanishing probability of converting to axions via a see-saw-like mechanism after travelling a distance $L$. The probability is given by \cite{Csaki:2001yk, Deffayet:2001pc}:
\be
P_{\gamma \rightarrow \phi} = \sin(2\theta)^2 \sin\left(\frac{\Delta}{\cos(2\theta)}\right)^2,
\ee
where the parameters in the previous equation are given by $\Delta = m_{\rm eff}^2 L / 4 \omega$, $\tan (2\theta)=2B\omega/(M m_{\rm eff}^2)$. Here, $B$ is the strength of the magnetic field, while $\omega=2\pi f$ the frequency of the photons and $m_{\rm eff}^2=|m_\phi^2-\omega_P^2|$, where $\omega_P^2=4\pi^2 \alpha n_e/m_e$ is the plasma frequency of the medium related to the effective mass of the photons and  $m_\phi$ is the axion mass.

This probability of converting photons to axions means that the photon number is not conserved, hence the observed luminosity distance, $d_{L}^{\rm EM}(z)$, is different to the ``bare'' one, $d_{L}^{\rm bare}(z)$, which corresponds to a model where the photon number is conserved and can be computed using \autoref{eq:std_DDR}. Since we can only detect those photons along the line of sight, the observed and bare luminosity distances are related by a factor $\mathcal{P}(z)$ such that \cite{Avgoustidis:2010ju}
\be \label{eq:em_deviation}
d_{L}^{\rm EM}(z)= \frac{d_{L}^{\rm bare}(z)}{\sqrt{\mathcal{P}(z)}}\, .
\ee 
The redshift evolution of the function $\mathcal{P}(z)$ depends on the type of intervening magnetic field responsible for the photon decay. Following \cite{Avgoustidis:2010ju}, we distinguish here between incoherent (inc) and coherent (coh) magnetic fields, leading to different redshift trends for the $\mathcal{P}(z)$ function:
\begin{equation}
\mathcal{P}_{\rm{inc}}(z)=A+(1-A) \exp\left(-\frac32 \frac{H_0}{c}r(z) ~\xi_0\right)\, ,\label{eq:Pinc}
\end{equation}
\begin{equation}
\mathcal{P}_{\rm{coh}}(z)=A+(1-A) \exp\left(-\frac{H(z)-H_0}{\Omega_{\rm m} H_0} ~ \xi_0\right)\, , \label{eq:Pcoh}
\end{equation}
where $\Omega_{\rm m}$ is the energy density of matter at $z=0$. The factor $A$ sets the amplitude of the deviation from the standard DDR, and it can be expressed in terms of the initial flux of the axions and photons at some initial redshift $z_I$, denoted by $I_\phi(z_I)$ and $I_\gamma(z_I)$ respectively, as \cite{Avgoustidis:2010ju}
\begin{equation}\label{eq:A}
A=\frac{2}{3}\left(1+\frac{I_\phi(z_I)}{I_\gamma(z_I)}\right)\, ,
\end{equation}
and the parameter $\xi_0$ is related to the transition probability $P_{\gamma \rightarrow \phi}$ of each domain of length $L$ through
\begin{equation}\label{eq:xi0}
    \xi_0=\frac{c}{H_0}\frac{P_{\gamma \rightarrow \phi}}{L}\, .
\end{equation}
Since we expect the photons to travel through several domains of intergalactic magnetic fields with coherence of at least $\sim 50 \textrm{Mpc}$, then \autoref{eq:Pinc} and \autoref{eq:Pcoh}  can be considered as both an average over several domains and frequencies of the photons. Moreover, we may make a heuristic argument that the transition probability should be of the order of a few percent, which then implies from \autoref{eq:xi0} that $\xi_0=\mathcal{O}(1)$. Hence, throughout the rest of this paper, we will assume that $\xi_0=1$.

Assuming the angular diameter distance $d_A(z)$ is not affected, we can then define the parameter $\eta_{\rm EM}(z)$ which characterises the deviation from the DDR:
\begin{equation}\label{eq:eta_axions}
\eta_{\rm EM}(z)\equiv\frac{d^{\rm EM}_L(z)}{d^{\rm bare}_L(z)}=\frac{d^{\rm EM}_L(z)}{(1+z)^2 d_A(z)} = \frac{1}{\sqrt{\mathcal{P}(z)}},
\end{equation} 
where $\mathcal{P}(z)$ is given by \autoref{eq:Pinc} and \autoref{eq:Pcoh} in the incoherent and coherent regimes respectively. Previous literature investigating departures from the DDR usually makes use of a simple parameterisation (see e.g. \cite{Avgoustidis2009,Avgoustidis:2010ju})
\begin{equation}\label{eq:eps0par}
    d_{L}^{\rm EM}=(1+z)^{\epsilon_0}d_{L}^{\rm bare}(z)\, ,
\end{equation}
which yields
\begin{equation}\label{eq:eta_eps0}
    \eta_{\rm EM}(z) = (1+z)^{\epsilon_0}\, .
\end{equation}
Therefore, one can compare \autoref{eq:eta_axions} and \autoref{eq:eta_eps0} to map current constraints on $\epsilon_0$. 
From \autoref{eq:A} we see that if $A>1$, then this implies that at early times the intensity of the axions satisfies $I_\gamma(z_I) < 2 I_\phi(z_I)$. Since we roughly expect the intensity of the particles to be proportional to their number $n$, then  this also implies that approximately $n_\gamma(z_I) < 2 n_\phi(z_I)$. Mapping current constraints on $\eta_{\textrm{EM}}(z)$ \cite{EUCLIDWP10} to the $\mathcal{P}(z)$ function through \autoref{eq:eta_axions}, we obtain, in the coherent regime, $A=\mathcal{O}(1)$, which implies that $n_\gamma(z_I)\simeq 1.881 n_\phi(z_I)$ and it is consistent with the fact that photons have two polarisations while axions have only one, and at at early times they are all in a thermal equilibrium.

\begin{figure}[!t]
	\centering
	\includegraphics[width=0.8\textwidth]{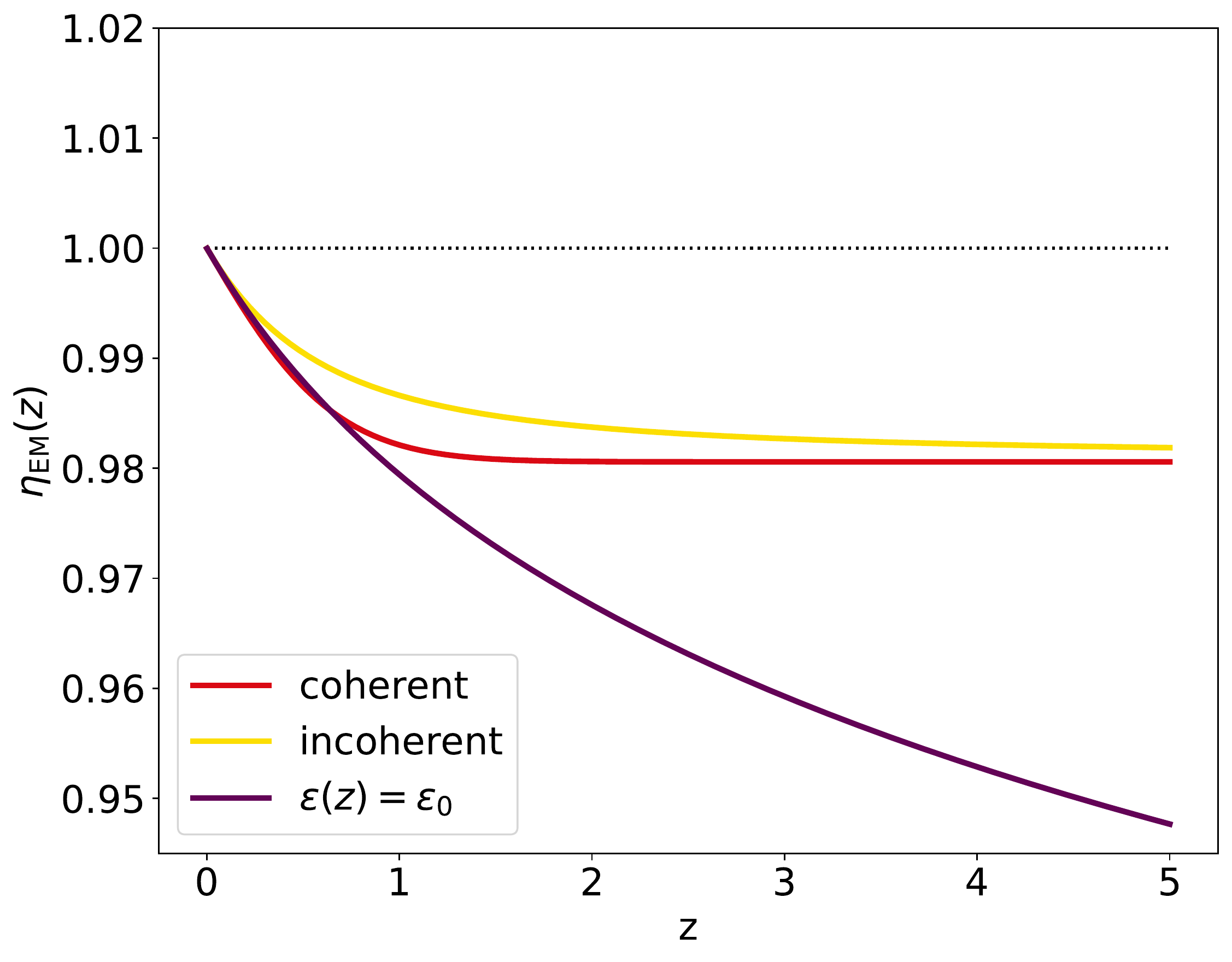}
	\caption{A comparison of the incoherent and coherent models, given by \autoref{eq:Pinc} and \autoref{eq:Pcoh} respectively, versus the phenomenological expansion $\epsilon(z)=\epsilon_0$. We assumed $\Omega_{m0}=0.315$ and $\epsilon_0=-0.03$.}\label{fig:etaz}
\end{figure}

In \autoref{fig:etaz} we show a comparison of the duality parameter $\eta_{\textrm{EM}}(z)$ for the incoherent and coherent axion models, given by \autoref{eq:Pinc} and \autoref{eq:Pcoh} respectively, versus the phenomenological parameterisation of \autoref{eq:eta_eps0}. We assumed $\Omega_{\rm m}=0.315$ and $\epsilon_0=-0.03$. Note that at $z=0$, all models have $\eta_{\textrm{EM}}(z=0)=1$. Since the conversion of photons to axions is an integrated effect along the line of sight, at small distances away from the observer there are very few, if any, magnetic domains, and therefore the distance duality relation holds. It is worth to point out that models of the kind investigated here might also lead to time variation of fundamental constant such as the fine structure constant $\alpha$ (see e.g. \cite{Hees:2014lfa}).

\section{Forecast DDR constraints: the impact of standard sirens\label{sec:LCDM}}

In this paper we are interested in quantifying the constraints on the possible deviations from DDR due to the mechanisms described in Section \ref{sec:theory}. The crucial observations needed to constrain these effects are produced by SNIa surveys, which will provide measurements of the distance modulus $\mu(z)$, connected to the luminosity distance through
\begin{equation}
    \mu(z)\equiv m(z)-M_0=5\log_{10}{d_{L}^{\rm EM}(z)}+25\,,
\end{equation}
where $m$ is the apparent magnitude of the SNIa at redshift $z$ and $M_0$ its intrinsic magnitude. Such measurements are sensitive to the decay of photons through \autoref{eq:em_deviation} and can therefore place constraints on the parameters that govern the coupling of photons to axions, $A$ and $\xi_0$. However, it has been shown \cite{EUCLIDWP10} that using the information from SNIa surveys alone leads to strong degeneracies between the DDR parameters and $\Omega_{\rm m}$, limiting the constraining power of this observable (more details on this degeneracy are shown in \autoref{subsec:results:constraints} below).

For this reason, it is common to combine SNIa with BAO data; the latter are able to provide constraints on combinations of the angular diameter distance $d_A(z)$, the Hubble parameter $H(z)$ and the sound horizon at the dragging epoch $r_d$. These measurements are not sensitive to the deviation from standard DDR that we consider and can therefore be used to break the degeneracies and increase the constraining power of the data on $A$.

However, the BAO data come with their own issues. A fiducial cosmology must be assumed to obtain distances from the measured angular scale on the sky, thus possibly inducing some model bias into the data. Further uncertainties are introduced by the fact that nonlinear effects damp and modify the locations of the BAO in the galaxy power spectrum, thus possibly introducing systematic errors in the estimation of the inferred cosmological parameters (see e.g. \cite{Angulo:2007fw}). Several techniques have been developed to standardise BAO distance measurements; however most of them rely on modelling of nonlinear scales, which is not trivial if one abandons the $\Lambda$CDM model for extended theories. Alternatively one can rely on observables not affected by such nonlinear effects \cite{Anselmi:2017cuq,Anselmi:2017zss}, paying the price of a reduced constraining power.

Given these caveats, it would be useful to have an extra observable to use alongside the routinely employed SNIa and BAO; such observables would need to be able to probe the cosmological parameters without suffering from the degeneracy with DDR deviations that SNIa exhibit. Recent detections of gravitational waves (e.g. \cite{LIGOScientific:2018mvr, Abbott:2020uma, LIGO:2004.08342v2}) have shown that these observations can provide a new way of testing the fundamental physical mechanisms at play in the Universe. Gravitational waves are the propagation of perturbations in the tensor sector, which, in GR and in vacuum propagation satisfy
\begin{equation}\label{eq:GRprop}
h''_\mathcal{A}(\tau,k)+2\mathcal{H}h'_\mathcal{A}(\tau,k)+k^2h_\mathcal{A}(\tau,k)=0\, \
\end{equation}
with $h_\mathcal{A}(\tau,k)$ the Fourier modes of the GW amplitude, the prime representing the derivative with respect to conformal time $d\tau = dt/a(t)$, $\mathcal{H}$ the conformal Hubble parameter ($\mathcal{H}= aH$) and
the index $\mathcal{A}=+,\times$ running over the two polarisations.

It can be shown \cite{Maggiore2007} that $h_\mathcal{A}$ scales with the luminosity distance as
\begin{equation}\label{eq:GW_dl_relation}
    h_\mathcal{A}\propto\frac{1}{d^{\rm GW}_L(z)}\, ,
\end{equation}
and therefore distance measurements can be obtained by observing gravitational waves from merger events. If the redshift of the event is measured by observing an electromagnetic counterpart, we can construct a Hubble diagram using these as standard sirens.

The photon--axion coupling we consider in this work does not affect the luminosity distance measured through GW observations. These therefore probe the bare luminosity distance, assuming that no other physical mechanism is leading to deviations from the GW propagation predicted by GR, and in \autoref{eq:GW_dl_relation} $d_L^{\rm GW}(z)=d_{L}^{\rm bare}(z)$. This implies that, as with the BAO data, the observations  of standard sirens by future surveys can be used in combination with SNIa to constrain deviations from DDR. We therefore focus on these three observables: SNIa, BAO and GW, using them to quantify our future ability to constrain deviations from the DDR. We create simulated data for: 
\begin{itemize}
    \item an SNIa survey based on what will be achievable with the Legacy Survey of Space and Time (LSST), performed by the Vera C. Rubin Observatory. LSST will survey approximately 18,000 square degrees of the sky and conservative estimates predict observations of 10,000 SNIa up to $z\approx1$ \cite{Abell:2009aa}. We provide more details of the LSST data we simulate in Appendix \ref{sec:snmock}.
    \item a BAO survey based on forecast data for the Dark Energy Spectroscopic Instrument (DESI), a spectroscopic galaxy survey expected to be fully operational by the end of 2020  \cite{DESI2016}. We provide more details of the DESI data we simulate in Appendix \ref{sec:baomock}.
    \item GW data expected from the proposed Einstein Telescope (ET) \cite{Maggiore:2019uih}, a future third-generation terrestrial gravitational wave observatory. We consider here future observations of binary neutron star (BNS) mergers; such events could provide a corresponding electromagnetic observation, allowing a redshift measurement, and in this paper we assume that a counterpart will be available for $N_{\rm GW}=1000$ observations performed by ET\footnote{Such an assumption could be seen as optimistic. If the number of events with an electromagnetic counterpart is reduced, one could infer the redshift of the mergers with alternative methods, paying the price of a larger uncertainty on the redshift estimation (see e.g. \cite{Chen:2017rfc})}. We follow the specifications and the noise calculation presented in \cite{Sathyaprakash:2009xt,Zhao:2010sz, Du:2018tia}. A detailed description of the steps and assumptions taken to obtain the simulated dataset for GW is shown in Appendix \ref{sec:gwmock}.
\end{itemize}

Note that the ET as an experiment is still in the proposal stage, as opposed to DESI, which has already seen first light, and LSST, which is currently under construction. This means that there is likely to be a $\sim 10$ year gap between the final data releases from LSST and DESI and the first results from ET. Nevertheless, forecasting the constraints that all three will jointly provide is still an interesting and useful endeavour.

Following the specifications for these experiments we assume a fiducial cosmology where no deviation from DDR is present ($A=1$), and we take the fiducial values for the standard cosmological parameters $\Omega_{\rm m}=0.314$, $H_0=67.36$ km s$^{-1}$ Mpc$^{-1}$. Throughout this paper we assume a vanishing contribution to the total energy density from curvature ($\Omega_k=0$) and we assume that the late time expansion of the Universe is dominated by a cosmological constant $\Lambda$ with energy density $\Omega_\Lambda=1-\Omega_{\rm m}$. As the data we consider only probe the low redshift regime, we consider the contributions from the radiation energy density to be negligible.

We forecast the constraining power of these surveys, implementing a new likelihood module for the publicly available MCMC sampler \texttt{Cobaya} \cite{Torrado:2020dgo}, obtaining the theoretical prediction for $d_A(z)$ from \texttt{CAMB} \cite{Lewis:1999bs,Howlett:2012mh} for each point in the parameter space, and computing the luminosity distances observed by SNIa ($d_L^{\rm EM}$) and GW ($d_L^{\rm GW}$) using \autoref{eq:em_deviation} and \autoref{eq:dl} respectively. We sample the standard cosmological parameters $\Omega_{\rm m}$ and $H_0$, alongside the DDR parameter $A$, imposing flat priors on them. We obtained our constraints for both the incoherent and coherent axion models, finding no significant difference in the results for each. We thus choose to present only the incoherent model results.

When fitting the simulated SNIa data we make use of the likelihood described in Appendix C of \cite{Conley:2011ku}, which takes into account the complete degeneracy of $H_0$ and $M_0$ that this probe suffers from, marginalising them out.

\subsection{Results} \label{subsec:results:constraints}

Using the mock datasets introduced above, where no deviation from DDR occurs, we aim to forecast the constraints that will be achieved in the future using the three types of observations we consider here. We firstly focus on the results obtained using SNIa observations alone. As we already discussed, these observations are not sensitive to $H_0$ and $M_0$ individually, but rather only to their combination. 

\begin{figure}
    \centering 
    \includegraphics[width=0.5\textwidth]{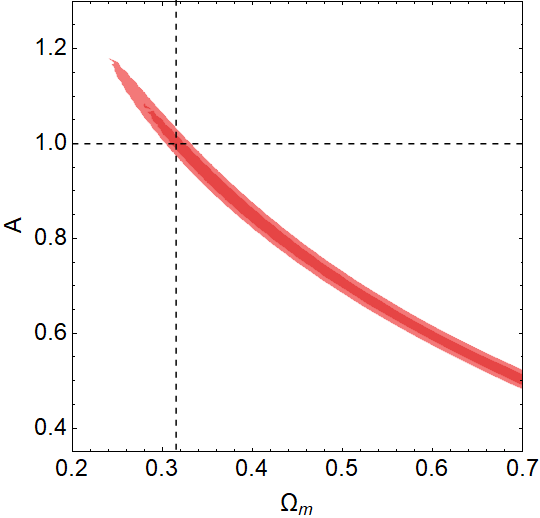} 
    \caption{2D constraints on the matter energy density $\Omega_{\rm m}$ and the DDR violation parameter $A$ obtained analysing the LSST mock dataset for SNIa.}
    \label{fig:SNdeg}
\end{figure}

However, another strong degeneracy appears when we try to constrain deviations from the standard DDR. As we show in \autoref{fig:SNdeg}, the parameter $A$ is strongly degenerate with $\Omega_{\rm m}$, and a variation of $A$ allows the theoretical predictions we obtain to be compatible with the dataset for  extreme values of the matter energy density. Notice that to obtain the results of \autoref{fig:SNdeg}, we relied on a grid sampling of the bi-dimensional parameter space, rather than on MCMC method based on the Metropolis--Hastings (MH) algorithm \cite{Hastings1970}, such as that implemented in \texttt{Cobaya}, that we use for the rest of the results. This is necessary due to the degeneracy itself, which makes the MCMC struggle to reconstruct the posterior distribution. When sampling these parameters using the LSST mock dataset alone, the MH struggles to explore the full line of degeneracy between $A$ and $\Omega_{\rm m}$, instead finding false peaks in the posterior distribution which it is unable to move away from. 

As a further check, we exploited the method of nested sampling, using the \texttt{PolyChordLite} code \cite{Handley2015a, Handley2015b} implemented in \texttt{Cobaya}. This enabled us to properly explore the full extent of the degeneracy, as nested sampling is much better suited to sampling multi-modal and other complicated distributions than MH, and also allowed us to recover the results obtained with the grid approach. We were further able to show that the addition of the BAO dataset is sufficient to break the degeneracy between $\Omega_{\rm m}$ and $A$. We therefore urge caution when investigating degenerate models with simple sampling methods such as MH, and stress that checks with different sampling methods are always beneficial. Such a result highlights the necessity of using other observations which, unlike SNIa, are not sensitive to the parameter $A$ and are therefore able to break this degeneracy by measuring $\Omega_{\rm m}$.

\begin{figure}
    \centering
    \includegraphics[width=0.8\textwidth]{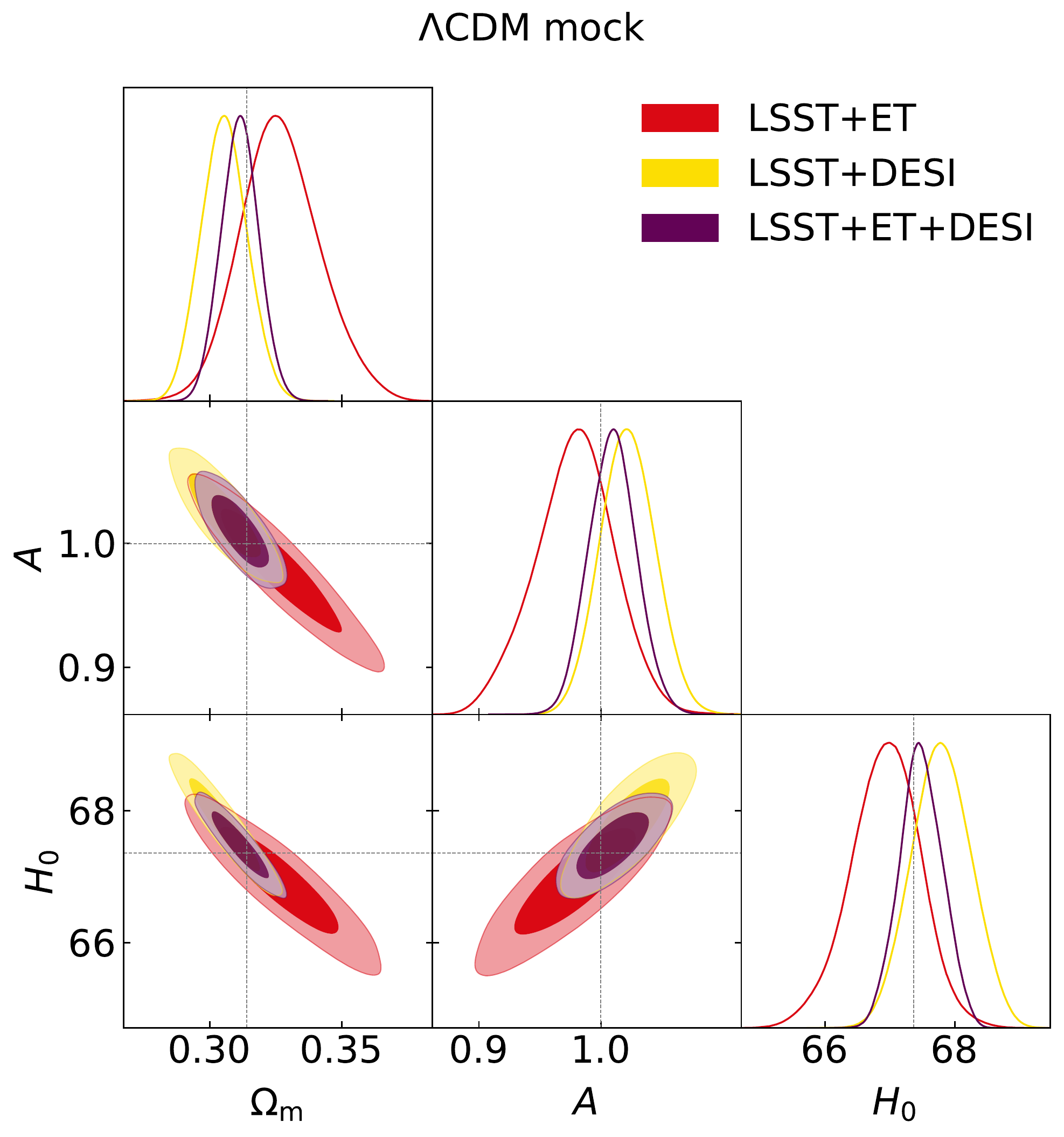} \\
    \caption{Constraints on $H_0$, $\Omega_m$ and the photon--axion decay model parameter $A$ for electromagnetic DDR breaking, using a mock obtained with a $\Lambda$CDM cosmology. The combinations of LSST+ET, LSST+DESI and LSST+ET+DESI are shown in red, yellow and purple respectively.}
    \label{fig:axionpars_LCDMmock}
\end{figure}

We therefore now focus on the combination of SNIa with BAO and GW observations. Our constraints on the cosmological parameters are reported in \autoref{tab:resLCDM}, while we show the marginalised posterior distributions and the two dimensional contours in \autoref{fig:axionpars_LCDMmock}, combining LSST first with ET and DESI separately and then all together. The $68$\% confidence intervals of the constraints are all compatible with the $\Lambda$CDM fiducial cosmology, shown by the dashed lines.

These results show how a combination of SNIa and GW observation from the ET would be competitive with the combination of SNIa and BAO, with the constraints that improve only by a factor of $\approx1.4$ for the latter. The use of standard sirens is therefore able to break the degeneracy between $A$ and $\Omega_{\rm m}$ and to help in constraining deviations from DDR, thus allowing us to test whether or not the BAO results are affected by the possible issues we described above. Nevertheless, it is possible to notice in \autoref{fig:axionpars_LCDMmock} how the $A-\Omega_{\rm m}$ degeneracy is not completely broken by the use of ET data, with the $\Omega_{\rm m}$ posterior moving towards high values because of this effect.

Combining all datasets together the constraints on DDR violations are improved with respect to the LSST+ET and LSST+DESI cases, and we achieve a $2$\% constraint on $A$.

\begin{table}[!htbp]
\begin{center}
\begin{tabular}{l|c|c|c} 
\hline
                   & LSST+ET          & LSST+DESI          & LSST+ET+DESI \\
\hline

$H_0$              & $66.87\pm 0.54$  & $67.75\pm 0.47$    & $67.44\pm 0.36$\\
\hline
$\Omega_{\rm m,0}$ & $0.328\pm 0.015$ & $0.3056\pm 0.0090$ & $0.3116\pm 0.0077$\\
\hline
$A$                & $0.977\pm 0.033$ & $1.023\pm 0.023$   & $1.010\pm 0.020$\\
\hline
\hline 
\end{tabular}
\caption{Mean values and marginalised $68\%$ confidence level errors for $H_0$, $\Omega_{\rm m}$ and $A$ for the three combinations of mock datasets considered.}\label{tab:resLCDM}
\end{center}
\end{table}

\section{Modified Gravity effects on the luminosity distance}\label{sec:MG}

In \autoref{sec:LCDM} we have shown how the use of GW observations alongside SNIa allows us to obtain results competitive with the use of BAO on violations of the DDR, while at the same time avoiding the possible assumptions underlying the determination of the BAO data. However, several effects can alter the propagation of GW leading to $d_L^{\rm GW}(z)\neq d_L^{\rm bare}(z)$. If such effects are not properly taken into account, the analysis we proposed above can lead to inaccurate results, with a bias introduced on the estimation of cosmological parameters.

In order to show this possible setback in the use of GW, in this paper we focus on theories alternative to GR, such as theories that generalise the Einstein--Hilbert action by adding non-minimally coupled scalar fields or higher order covariant terms; in such cases, we expect modifications to the terms of \autoref{eq:GRprop}. Changes to the $k^2h_\mathcal{A}$ term cause the speed of propagation of GW ($c_T$) to vary and are therefore extremely constrained by the observations of the event GW170817 \cite{BNSmerger}, which determined the relative difference between $c_T$ and the speed of light to be $\mathcal{O}(10^{-15})$. However, such constraints are obtained for $z\lesssim0.1$, thus in principle a time-varying speed of the gravitational wave propagation could be allowed at higher redshifts. Changes to the friction term of GW propagation are also not excluded, so overall we can consider the modified propagation equation of the form \cite{Belgacem:2017ihm, Belgacem:2018lbp} 

\begin{equation}
h''_\mathcal{A}(\tau,k)+2\mathcal{H}[1-\delta(\tau)]h'_\mathcal{A}(\tau,k)+k^2c_T(\tau)^2h_\mathcal{A}(\tau,k)=0,\, \
\end{equation}
where $\delta(\tau)$ parameterises deviations from GR and is assumed to be scale independent. It can be shown that such modifications to the GW propagation leads to a departure of the gravitational wave luminosity distance from $d_L^{\rm bare}(z)$ \citep{Belgacem:2017ihm,Belgacem:2019pkk},
\begin{equation}\label{eq:gw_deviation}
    d_L^{\rm GW}(z)= \sqrt{\frac{c_T(z)}{c_T(z=0)}}\exp{\left[-\int_0^z{\frac{\delta(z')}{1+z'}dz'}\right]}(1+z)\int_0^z{\frac{c_T(z')}{H(z')}dz}\, ,
\end{equation}
which reduces to the standard luminosity distance for $\delta(z)=0$ and $c_T(z)=c$. In what follows we assume that the bound on $c_T$ provided by GW170817 holds at all redshifts, and therefore, setting $c_T(z)=c$, the previous equation reduces to
\begin{equation}\label{eq:gw_deviation1}
    d_L^{\rm GW}(z)= \exp{\left[-\int_0^z{\frac{\delta(z')}{1+z'}dz'}\right]}(1+z)\int_0^z{\frac{c}{H(z')}dz}= \exp{\left[-\int_0^z{\frac{\delta(z')}{1+z'}dz'}\right]}d_L^{\rm bare}(z)\, .
\end{equation}

To connect this expression to non-standard theories of gravity, we can use the relation between $\delta(\tau)$ and a time-varying effective Planck mass\footnote{As noted by \cite{Belgacem:2019pkk}, this relation is not universally true for every theory of modified gravity and so a non-zero $\delta(\tau)$ should not be immediately associated with a time-varying Planck mass. However, for the purpose of our investigation, it is a suitable choice.}, $M_{\mathrm{eff}}$ \cite{Belgacem:2019pkk},
\begin{equation}
\delta(\tau) = -\frac{d \ln M_{\mathrm{eff}}}{d \ln a}, \label{eq:planckmass}
\end{equation}
which means we can rewrite \autoref{eq:gw_deviation1} as
\begin{equation}
\frac{d_L^{\rm GW}(z)} {d_{L}^{\rm bare}(z)} =\frac{M_{\mathrm{eff}}(0)}{M_{\mathrm{eff}}(z)}\, .\label{eq:dLGeff}
\end{equation}
Since the effective Planck mass is related to the effective Newton's constant, $G_{\rm eff} \propto 1/M_{\rm eff}^2$, we can recast this as
\begin{equation}\label{eq:GWMG}
\frac{d_L^{\rm GW}(z)} {d_L^{\rm bare}(z)}=\sqrt{\frac{G_{\mathrm{eff}}(z)}{G_{\mathrm{eff}}(0)}}\, .
\end{equation}

An example of a non-standard theory that results in a time-varying Newton's constant can be found by examining the well-known Horndeski action, which describes the most general four dimensional Lorentz invariant scalar-tensor theory that produces second-order equations of motion\footnote{We note that it was very recently shown how a combination of SNIa and GW events would be able to probe dark energy fluctuations and a possible running of the Planck mass in the context of Degenerate Higher-Order Scalar-Tensor (DHOST) theories \cite{Garoffolo:2020vtd}.} \cite{Horndeski1974}. The action is given by
\begin{equation}
    S = \int d^4 x \sqrt{-g} \left[ \sum_{i=2}^{5} \mathcal{L}_i + \mathcal{L}_m\right],
\end{equation}
where the Lagrangian densities $\mathcal{L}_i$ are
\begin{align}
 \mathcal{L}_2 &= G_2(\varphi, X), \\
 \mathcal{L}_3 &= G_3 (\varphi, X) \Box \varphi, \\
 \mathcal{L}_4 &= G_4(\varphi, X)R + G_{4X}(\varphi, X) [(\Box \varphi)^2 - (\nabla_\mu \nabla_\mu \varphi)^2], \\
 \mathcal{L}_5 &= G_5(\varphi, X) G_{\mu\nu} \nabla^\mu \nabla^\nu \varphi - \frac{1}{6}G_{5X}(\varphi, X) [(\Box \varphi)^3- 3\Box \varphi(\nabla_\mu \nabla_\nu \varphi)^2 + 2 (\nabla_\mu \nabla_\nu \varphi)^3],
\end{align}
where $X= -\frac{1}{2}\partial_\mu \varphi \partial^\mu \varphi$, $R$ is the Ricci scalar, $G_{\mu\nu}$ is the Einstein tensor, and $\varphi$ is the additional scalar field of the Horndeski theory. Minimally coupled matter fields are contained in $\mathcal{L}_m$. The action is simplified by the binary neutron star merger constraint on the tensor speed, which implies that $G_{4X} = G_5 \approx 0$. 

Theories with a surviving quartic Galileon term $G_4$ result in a time-varying Planck mass, $M(t) = M_P \sqrt{G_4(\varphi)}$, which corresponds to an effective Newton's constant \cite{Dalang:2019fma}
\begin{equation}
    G_{\rm eff}(t) = \frac{G_N}{G_4(\varphi)}.\label{eq:geffMp}
\end{equation}

Note that here we focus on the effect of a time-varying Planck mass and not on the effective Newton's constant $G_{\rm eff}(z,k)$ which has a $k$ dependence and manifests as the effective gravitational constant between two test masses. This $k$ dependence manifests itself for example in first-order perturbation theory of $f(R)$ and scalar-tensor models \cite{Tsujikawa:2007gd, Nesseris:2008mq,Nesseris:2009jf,Arjona:2018jhh,Arjona:2019rfn} and in generalised scalar-tensor models of the  $f(R,\varphi, X)$ type, where $X=-\frac12 \varphi_{,c}\varphi^{,c}$ is the kinetic term. In this case the effective Newton's constant $G_{\rm eff}(z,k)$ is then given by \cite{Tsujikawa:2007gd}
\be
G_{\rm eff}(z,k)=\frac{1}{F}\frac{f_{,X}+4\left(f_{,X} \frac{k^2}{a^2}\frac{F_{,R}}{F}+\frac{F_{,\varphi}^2}{F}\right)}{f_{,X}+3\left(f_{,X} \frac{k^2}{a^2}\frac{F_{,R}}{F}+\frac{F_{,\varphi}^2}{F}\right)},\label{eq:gefffR}
\ee
where $F=f'(R)=f_{,R}$ and $F_{,R}=F'(R)$. However, as mentioned earlier, we will not consider this $k$ dependence here, only assuming a time dependence of the effective Newton's constant, which implies only a time-varying Planck mass, as in \autoref{eq:geffMp}.

In theories where Newton's constant is associated with a time-varying Planck mass, the peak luminosity of SNIa will also exhibit a dependence on such time variation. This is due to the fact that the peak SNIa luminosity is  proportional to the mass of nickel synthesised in the supernova \cite{Gaztanaga:2001fh}, which is a fixed fraction of the Chandrasekhar mass $M_{\textrm{Ch}}$. The latter varies as $M_{\textrm{Ch}}\sim G_{\textrm{eff}}^{-3/2}$, and as a result the SNIa peak luminosity varies as $L\sim G_{\textrm{eff}}^{-3/2}$. Thus, the absolute magnitude of the SNIa will acquire a correction of the form \cite{Nesseris:2006jc}
\be
M(z) = M_0+\frac{15}{4} \log_{10} \left[\frac{G_{\textrm{eff}}(z)}{G_{\textrm{eff}}(0)}\right], \label{eq:M0eq}
\ee
where $G_{\textrm{eff}}(0)\equiv G_N$ is the current value of Newton's constant as measured in a Cavendish experiment in a laboratory setting. This equation implies that the distance modulus now also acquires an extra correction of the form 
\ba
    \mu(z)&\equiv& m(z)-M_0\nn\\
    &=&5\log_{10}{d_\textrm{L}^{\rm EM}(z)}+25-\frac{15}{4}\log_{10}\left(\frac{G_{\textrm{eff}}(z)}{G_{\textrm{eff}}(0)}\right)\,.\label{eq:SNMG}
\ea
Notice that the effect of MG on observables shows an interesting similarity with models in which the fine structure constant
$\alpha$ is allowed to vary in redshift. In such models there is also a dependency of the observables on the ratio 
of $\alpha$ taken at different redshifts (emission and observation). Such similarity arises from the fact that in MG theories falling into the class of scalar-tensor theories, the additional scalar degree of freedom produces a non-minimal coupling to the matter sector in the Einstein frame. This is similar to what happens in varying $\alpha$ models, where an additional scalar degree of freedom is coupled with the electromagnetic sector and impacts cosmological observables in a similar way \cite{Calabrese:2013lga}.

Despite assuming a modified theory of gravity, we still fix the background expansion history in our model to that of a flat \lcdm model, with fiducial parameters as discussed in \autoref{sec:LCDM}, and neglect possible deviations introduced by modifications of gravity. Such a choice is common in the investigation of modified gravity theories, see e.g. \cite{Aghanim:2018eyx}, and it arises from the tight constraints that current data place on most of these theories, making their background expansions almost identical to that of \lcdm\ ; as an example, current data constrain $f(R)$ theories, included in the Horndeski class described above, to a level in which any deviation from the standard behaviour needs to be in the perturbation sector, since the allowed parameter space produces a background that mimics that of \lcdm \cite{Lombriser:2014dua,2019arXiv190803430B,Desmond2020}.

We should point out that recent studies on the absolute magnitude dependency on $G_{\rm eff}$ have brought the relation of \autoref{eq:M0eq} and \autoref{eq:SNMG} into question. On the one hand it was proposed that even though the Chandrasekhar mass varies as $M_{\textrm{Ch}}\sim G_{\textrm{eff}}^{-3/2}$, there are other effects that cause the effective luminosity of the SNIa to scale as $L\sim  G_{\textrm{eff}}^{3/2}$, thus the $G_{\textrm{eff}}$ term in the absolute magnitude would have the opposite sign \cite{Wright:2017rsu}. On the other hand it was also suggested that the scaling of the Chandrasekhar mass in terms of $G_{\textrm{eff}}$ needs to be revised completely, and a relation $M_{\textrm{Ch}}\propto G_{\textrm{eff}}^{-1}$ should be considered, resulting in an absolute magnitude given by $\mathcal{M}(z) = \mathcal{M}_0+\frac{5}{2} \log_{10} \left[G_{\textrm{eff}}(z)/G_{\textrm{eff}}(0)\right]$ \cite{Dalang:2019fma}.
While the dependence on the specific parameterisation is important, in the current work we choose to use the standard expression as given by \autoref{eq:M0eq} since we are only interested in modelling the effects of the modified gravity model. Furthermore, we are only interested in demonstrating how big the effects of modified gravity can be, so we use the aforementioned parameterisation as it is representative of this class of models but also broad enough at the same time.

In order to parameterise the evolution of Newton's constant, we consider a parameterisation for  $G_{\textrm{eff}}$ of the form \cite{Nesseris:2017vor}
\be
\frac{G_{\textrm{eff}}(z)}{G_N}=1+g_a \left(\frac{z}{1+z}\right)^n-g_a \left(\frac{z}{1+z}\right)^{2n},\label{eq:Geff}
\ee
which is equal to unity at both early and late times, thus recovering the standard value of Newton's constant, but allowing it to vary in between. This parameterisation is actually a Taylor expansion of Newton’s constant $G_\textrm{eff}$ around $a=1$ and then expressed in terms of redshift $z$ via $a=\frac{1}{1+z}$. In Ref.~\cite{Nesseris:2017vor}, it was shown that this parameterisation can successfully fit the growth rate and CMB data, and due to its specific form it also allows us to avoid the stringent bounds imposed both at low redshift by Solar system tests \cite{Nesseris:2017vor} and at high redshifts by Big Bang nucleosynthesis (BBN) \cite{Bambi:2005fi}.

In this paper we want to assess how neglecting modified gravity effects in GW propagation might lead to a false detection of DDR violations. In order to quantify this effect we follow the same approach of \autoref{sec:LCDM} to generate mock datasets, this time using a fiducial cosmology that assumes the presence of modifications of gravity, but no violations of the standard DDR. Therefore, we set $A=1$, $n=2$ and generate two mocks with $g_a=0.1$ and $0.5$ that we call MG-low and MG-high respectively. The former value of $g_a$ is consistent data, while the latter is not, as CMB lensing severely limits the available parameter space \cite{Nesseris:2017vor}.
 Hence, the value $g_a=0.5$ is considered here as an extreme case, as an example that strongly highlights the degeneracy between the EM and GW sectors. Notice that while the BAO dataset is assumed to be unchanged, both the SNIa and GW observations are affected by the variation of $G_{\rm eff}$, according to \autoref{eq:SNMG} and \autoref{eq:GWMG} respectively. 

\subsection{MG as a contaminant for DDR constraints}\label{subsec:biasedDDRres}

We now analyse the datasets obtained with a fiducial cosmology which includes a variation of Newton's constant, but still following the same procedure used in \autoref{sec:LCDM}, i.e. assuming (wrongly) that the only mechanism that enables a deviation from the $\Lambda$CDM expectations is the decay of photons into axions. We expect that this assumption will lead to a biased estimation of cosmological and DDR parameters, and we therefore want to quantify the false detection of DDR violation that might arise when analysing data.

\begin{figure}
    \centering
    \begin{tabular}{cc}
    \includegraphics[width=0.49\columnwidth]{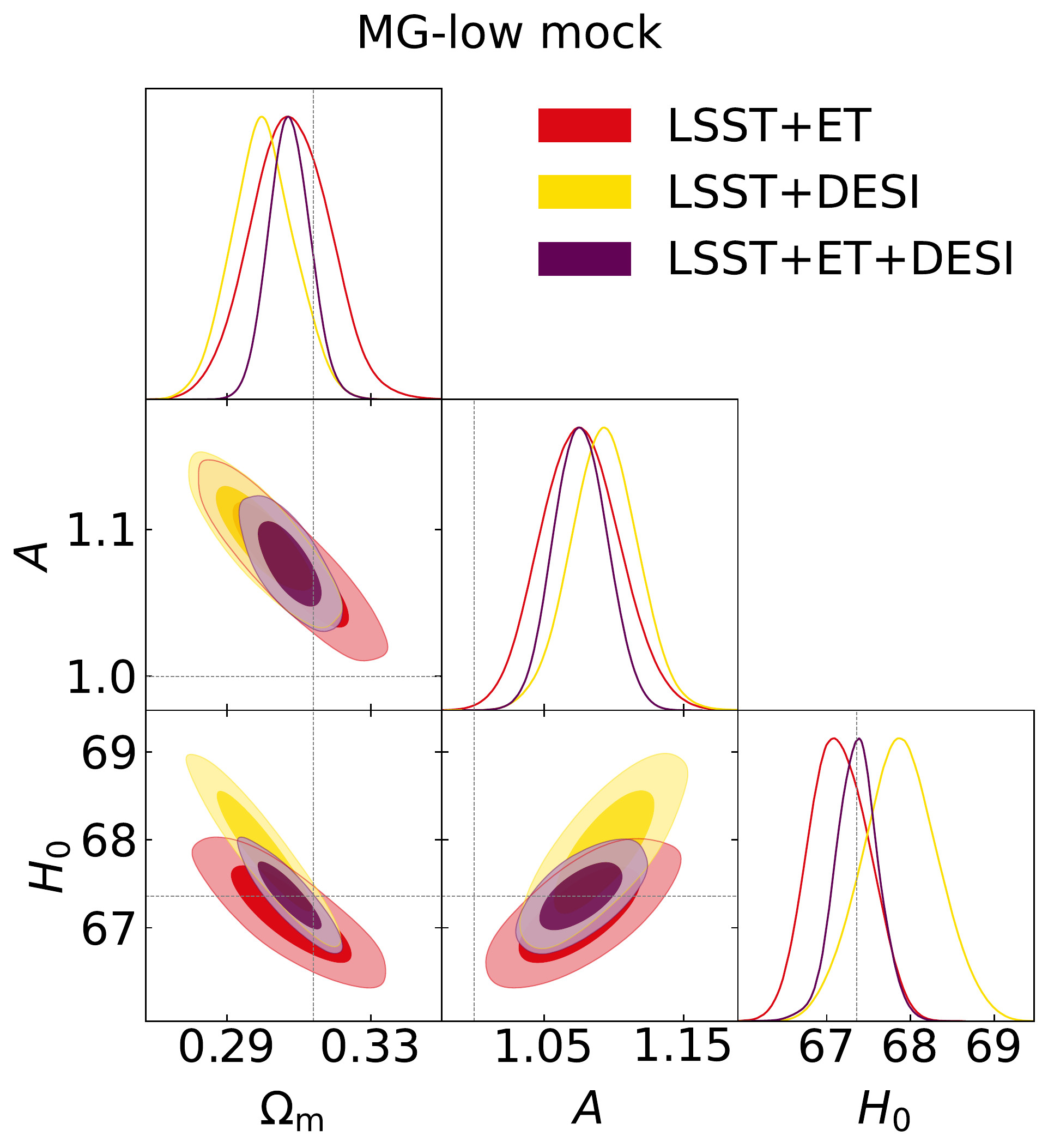} &
    \includegraphics[width=0.49\columnwidth]{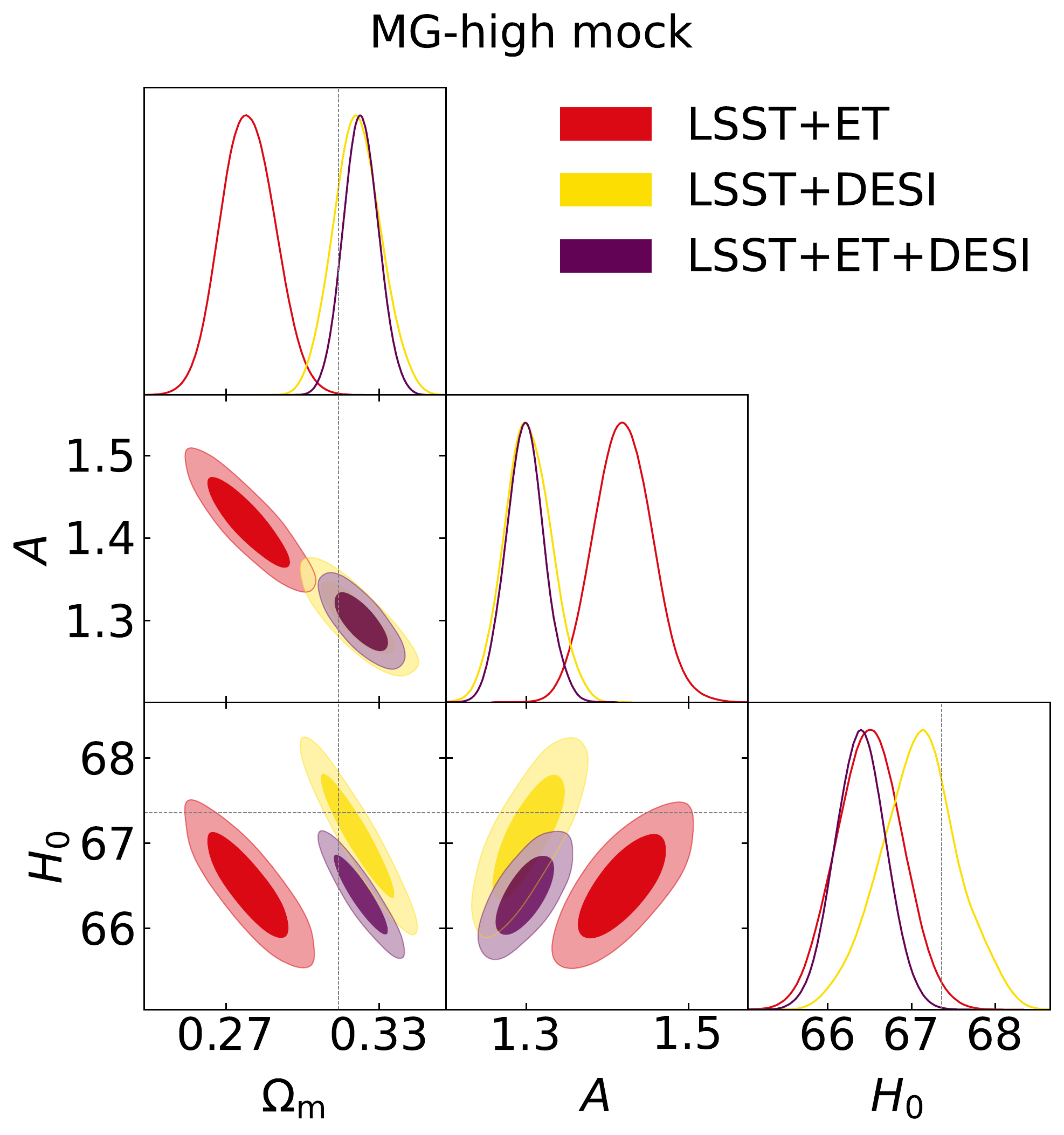}\\
    \end{tabular}
    \caption{Constraints on $H_0$, $\Omega_m$ and the photon--axion decay model parameter $A$ for electromagnetic DDR breaking, using a mock obtained with a MG cosmology (MG-low in the left panel, MG-high on the right). The combinations of LSST+ET, LSST+DESI and LSST+ET+DESI are shown in red, yellow and purple respectively.}
    \label{fig:resMGbias}
\end{figure}

We report the results of this analysis on both the MG-low and MG-high datasets in \autoref{tab:resMGbias} with the contours of the free parameters $\Omega_{\rm m}$, $H_0$ and $A$ shown in \autoref{fig:resMGbias}, with the $g_a$ parameter fixed to its standard value of $0$, which means that in our analysis $G_{\rm eff}(z)=G_N$. 
We find that the cosmological and DDR parameters are significantly biased in the MG-high case when the underlying fiducial MG cosmology is neglected, a bias that also appears in the MG-low case, albeit not as significant. For the latter, we find a false detection of DDR violations reaching $\approx4\sigma$, with the parameters $\Omega_{\rm m}$ and $H_0$ compatible with the fiducial cosmology within at most $2\sigma$. Instead, in the MG-high case, the LSST+DESI still recovers the fiducial values of $\Omega_{\rm m}$ and $H_0$ within $2\sigma$, but it now shows a striking false detection of $A\neq1$ at $\approx10\sigma$. Such a false detection is also present in the LSST+ET case, with similar significance, but in this case the cosmological parameters are also significantly biased away from their fiducial values ($\approx4\sigma$ for $\Omega_{\rm m}$ and $\approx2\sigma$ for $H_0$). Therefore, the two results are in tension with each other, a hint that despite both cases showing a strong detection of DDR violations, the mechanism considered is not sufficient to reproduce the observational data. We also report the combination of LSST+ET+DESI, even though such a combination is not very statistically sound, given the tension between the LSST+ET and LSST+DESI results shown in \autoref{fig:resMGbias}. We therefore make no further comment on this.

\begin{table}[!htbp]
\begin{center}
\begin{tabular}{l||c|c|c} 
\hline
            & \multicolumn{3}{c}{MG-low} \\
\hline
                   & LSST+ET          & LSST+DESI          & LSST+ET+DESI \\
\hline

$H_0$              & $67.16^{+0.34}_{-0.38}$ & $67.88\pm 0.45$ & $67.36\pm 0.26$  \\
\hline
$\Omega_{\rm m}$ & $0.308\pm 0.011$ & $0.3001\pm 0.0087$ & $0.3074\pm 0.0058$ \\
\hline
$A$                & $1.076\pm 0.028$ & $1.093\pm 0.024$ & $1.076\pm 0.019$ \\
\hline
\hline 
            & \multicolumn{3}{c}{MG-high} \\
\hline
                   & LSST+ET          & LSST+DESI          & LSST+ET+DESI \\
\hline

$H_0$              & $66.50\pm 0.40$ & $67.09\pm 0.47$ & $66.39\pm 0.30$  \\
\hline
$\Omega_{\rm m}$ & $0.279\pm 0.011$ & $0.3213\pm 0.0093$ & $0.3228\pm 0.0069$ \\
\hline
$A$                & $1.419\pm 0.036$ & $1.302\pm 0.029$& $1.299\pm 0.024$ \\
\hline
\hline
\end{tabular}
\caption{Mean values and marginalised $68\%$ confidence level errors for $H_0$, $\Omega_{\rm m}$ and $A$ for the three combinations of mock datasets considered, with the results for the MG-low and MG-high mocks shown separately.}\label{tab:resMGbias}
\end{center}
\end{table}

\subsection{Simultaneous MG and DDR constraints}\label{subsec:MGDDRres}

As a consequence of the false detection discussed in \autoref{subsec:biasedDDRres}, should a cosmological analysis find evidence for deviations from DDR, an analysis allowing for modifications of gravity should also be performed, as we have shown that its effects could be mistaken for violations of DDR.

Hence, we again analyse the MG-low and MG-high datasets, this time including $g_a$ as a free parameter, thus allowing for a redshift evolving Newton's constant. Notice that here we fix the parameter $n$ that enters into \autoref{eq:Geff} to its fiducial value of $2$, which is the minimum value allowed from Solar system tests \cite{Nesseris:2017vor}. Performing this analysis, we find a significant degeneracy between the $A$ and $g_a$ parameters, both in the LSST+DESI and LSST+ET combination, which make the posterior distributions very difficult to reconstruct through the MH approach we use in this work. It is easy to see the degeneracy in the LSST+DESI case, as the distance modulus we compare with SNIa data is given by \autoref{eq:SNMG} which can also be written, substituting \autoref{eq:dl}, as
\begin{equation}
    \mu(z)=5\log_{10}{d_\textrm{L}^{\rm bare}(z)}+25-\frac{5}{2}\log_{10}{\left[\mathcal{P}_{\rm inc}(z)\left(\frac{G_{\textrm{eff}}(z)}{G_{\textrm{eff}}(0)}\right)^{\frac{3}{2}}\right]}\,.\label{eq:SNMG_axion}
\end{equation}
It is clear how $A$, entering the expression of $\mathcal{P}_{\rm inc}(z)$, and $g_a$, ruling the deviations from $G_N$, can compensate each other to reproduce the mock data. As BAO are not sensitive to either of these parameters, this degeneracy will not be broken and the two parameters will be practically unconstrained.

One would expect the combination of LSST+ET not to suffer from this, as the GW are only sensitive to $g_a$ and should therefore break such a degeneracy. However, a variation of $g_a$ from its fiducial value can be partially compensated through a change in $\Omega_{\rm m}$ when analysing ET datasets, since ET has less constraining power than DESI on this parameter. This means that the degeneracy between $g_a$ and $A$ is also present in this combination and the parameters remain unconstrained.

\begin{figure}
    \centering
    \includegraphics[width=0.9\columnwidth]{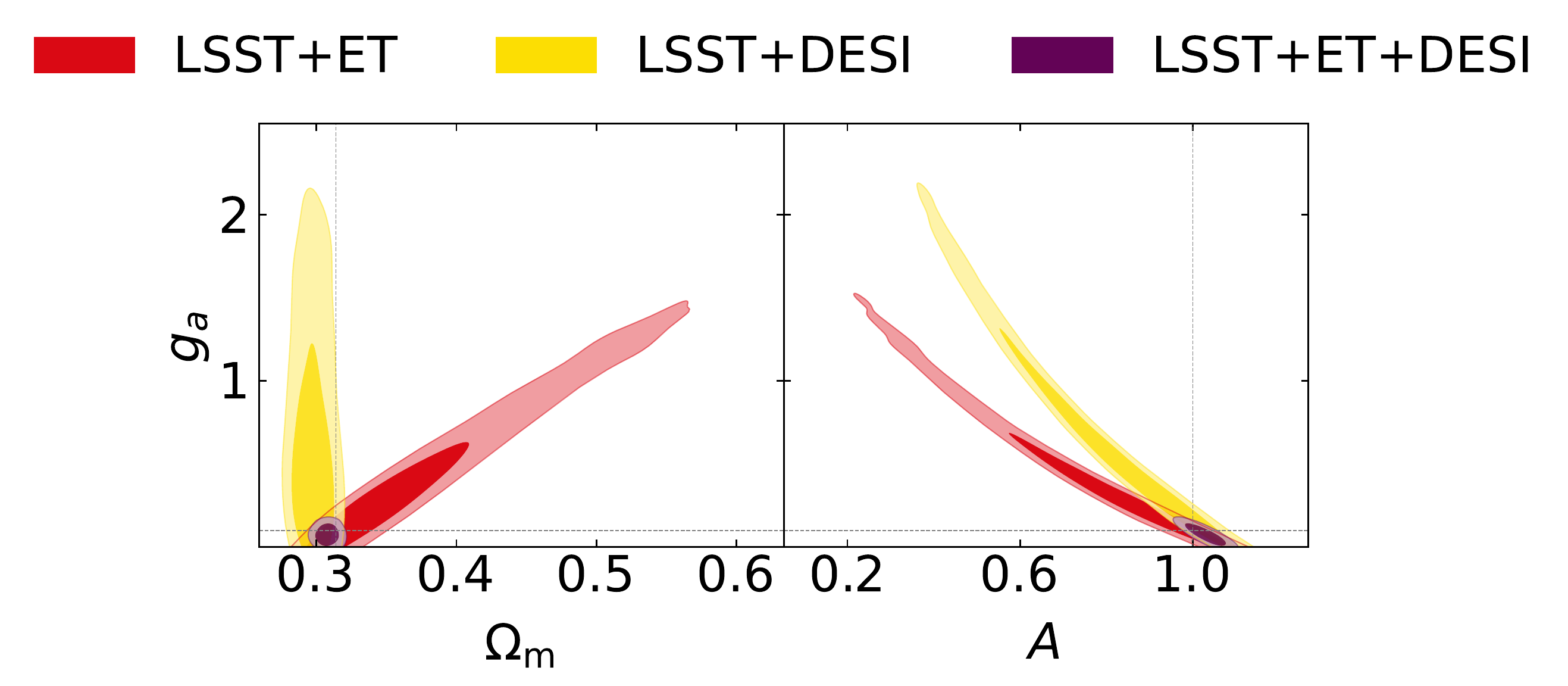}
    \caption{Constraints on $\Omega_m$, the photon--axion decay model parameter $A$ for electromagnetic DDR breaking, and the Newton's constant variation amplitude $g_a$. The results are obtained using a mock obtained with the MG-low fiducial cosmology. The combinations of LSST+ET, LSST+DESI and LSST+ET+DESI are shown in red, yellow and purple respectively.}
    \label{fig:resMGpars}
\end{figure}

We show this degeneracy in \autoref{fig:resMGpars} for the MG-low analysis; this result highlights the importance of combining these three observables when one wants to analyse both possible deviations from standard cosmology. Indeed, the addition of DESI to the LSST+ET combination breaks the degeneracy between $g_A$ and $\Omega_m$ in the analysis of ET data and, consequently, also breaks the degeneracy between $g_a$ and $A$, leading to strong constraints on both parameters. We report these constraints in \autoref{tab:resMGpars}, where we find that when combining LSST+ET+DESI we can constrain $A$ at the level of $\approx3\%$ even when the $g_a$ parameter is allowed to vary, with constraints of $\approx60\%$ and $\approx11\%$ for the latter in the MG-low and MG-high cases respectively.

\begin{table}[!htbp]
\begin{center}
\begin{tabular}{l||c|c|} 
\hline
            & MG-low & MG-high \\
\hline
                   &  LSST+ET+DESI      & LSST+ET+DESI \\
\hline

$H_0$              & $67.45\pm 0.26$    & $67.29\pm 0.33$   \\
\hline
$\Omega_{\rm m,0}$ & $0.3079\pm 0.0058$ & $0.3174^{+0.0067}_{-0.0075}$    \\
\hline
$A$                & $1.030\pm 0.033$   & $1.002\pm 0.033$    \\
\hline
$g_a$              & $0.077\pm 0.046$   & $0.482\pm 0.052$   \\
\hline
\hline 
\end{tabular}
\caption{Mean values and marginalised $68\%$ confidence level errors for $H_0$, $\Omega_{\rm m}$, $A$ and $g_a$ for the full combination of our mock datasets LSST+ET+DESI.}\label{tab:resMGpars}
\end{center}
\end{table}

\subsection{Impact of MG screening mechanisms}
Throughout this section we have included modified gravity effects in our analysis, implicitly assuming that for all the scales of interest these act in the same way. However, in modified gravity theories such as the well-known $f(R)$ \cite{Sotiriou:2008rp}, the equivalent scalar degree of freedom may develop an environment-dependent mass at small scales or when acting in a high density region. This dependence then makes the scalar field heavy enough that it screens the modifications of gravity, rendering them undetectable at those scales. 

A number of these screening mechanisms have been proposed, such as the chameleon screening \cite{Brax:2008hh} or the Vainshtein mechanism \cite{Vainshtein1973, Deffayet_2002} and it is clear that in the regimes of interest for our observables, i.e. the explosion of SNIa and the merging of binary neutron stars, one or other of these mechanisms might be active. The impact of screening on the observables we consider is rather difficult to predict, with the final result being strongly dependent on the specific mechanism at play. Focusing on GW propagation, and assuming that screening is active at the density and energy scales of the binary merger, the chameleon and Vainshtein mechanisms lead to different results; in the latter modifications are screened at the merger but still impact the propagation of the waves (see e.g. \cite{Jimenez:2015bwa}), in the former, any anomalous GW propagation might instead be completely screened away \cite{Dalang:2019fma, Dalang2019b}.

Here, we take a phenomenological approach and simply look at the case in which screening completely removes any modification to our observables, either in the explosion of the SNIa or in the propagation of GW from the BNS mergers. We do not consider the case in which the screening mechanism is active in both these astrophysical phenomena, as the resulting observations would be indistinguishable from the $\Lambda$CDM cosmology we assumed in \autoref{sec:LCDM}. 

\begin{figure}
    \centering
    \begin{tabular}{cc}
    \includegraphics[width=0.49\columnwidth]{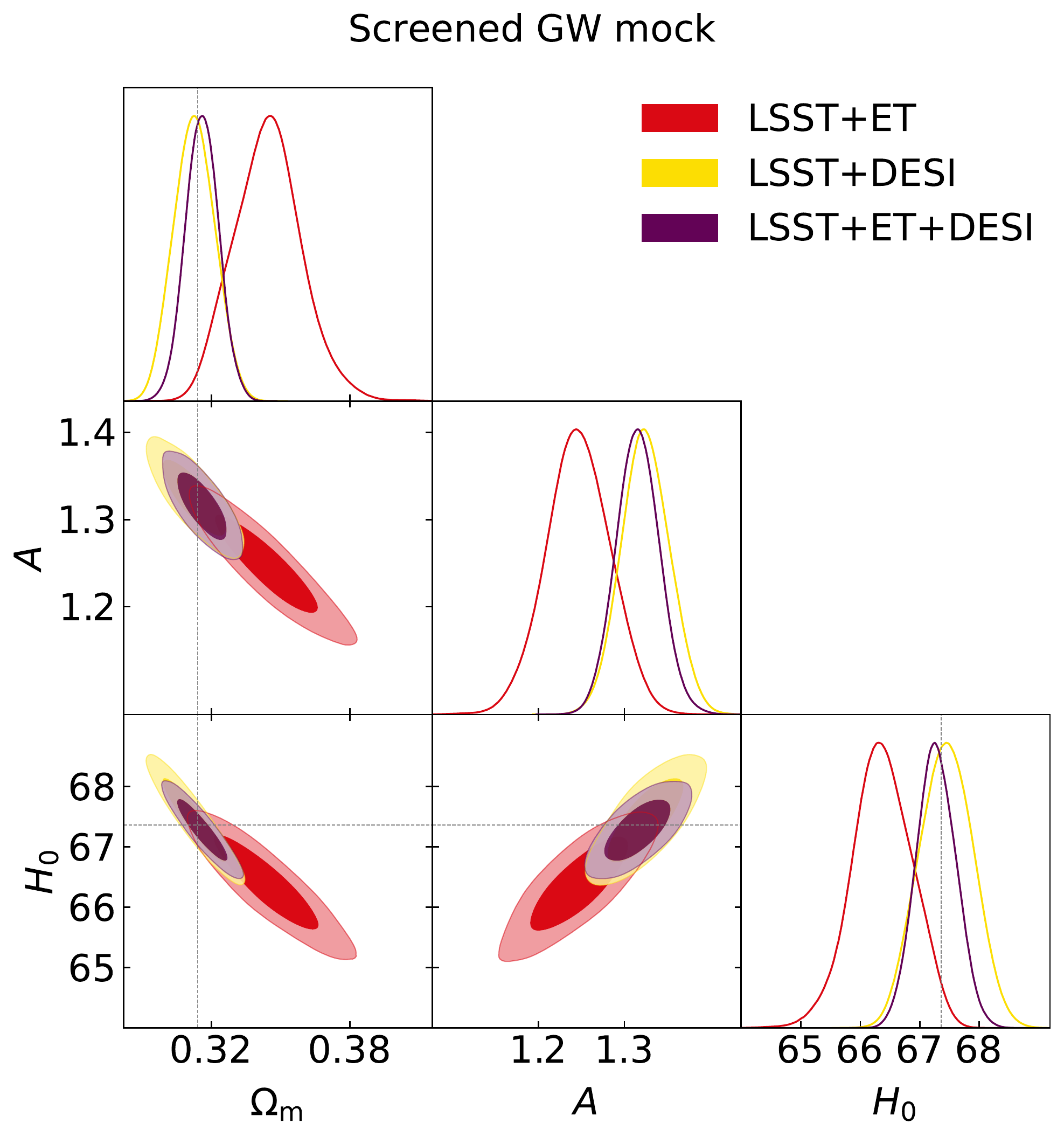} &
    \includegraphics[width=0.49\columnwidth]{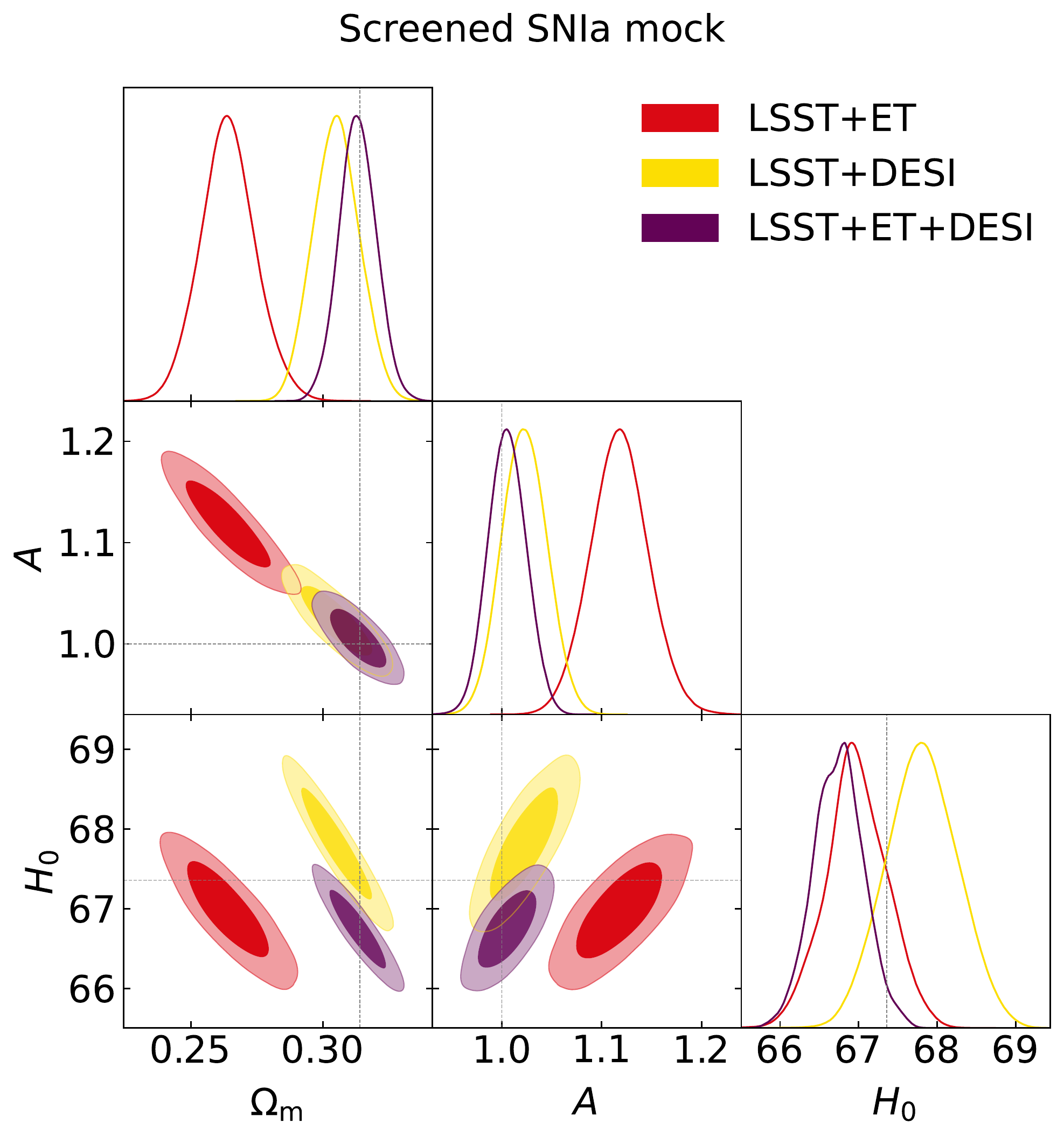}\\
    \end{tabular}
    \caption{Constraints on $H_0$, $\Omega_m$, and the photon--axion decay model parameter $A$ for electromagnetic DDR breaking. The results are obtained using a mock obtained the MG-high cosmology. In the left panel only the MG effects on GW propagation are screened, while in the right panel the screening only acts on SNIa. The combinations of LSST+ET, LSST+DESI and LSST+ET+DESI are shown in red, yellow and purple respectively.}
    \label{fig:screenres}
\end{figure}

In order to investigate this scenario we consider our MG-high settings, i.e. a departure from $G_N$ as parameterised in \autoref{eq:Geff} with $g_a=0.5$ and $n=2$, but removing the MG effects from either SNIa or GW. We analyse these datasets, again assuming that no MG effect is taking place, in order to assess the false detection of a DDR violation. The results are shown in \autoref{fig:screenres}, with the left panel obtained assuming GW are screened and SNIa are not, and the right assuming the opposite. 

We find that when GW are screened but SNIa are not, both the LSST+ET and LSST+DESI combinations find a false detection of DDR violation, as the unscreened SNIa dataset appears in all combinations. If instead SNIa are not affected by MG, while GW are, the MG effects only enter in the ET dataset. The combinations including this dataset are therefore biased away from the fiducial cosmology, meaning only the LSST+DESI combination correctly recovers the fiducial. If such a situation were to arise in real data, it could provide potential smoking gun for the effects of a modified gravity model with this type of screening behaviour.

\section{Machine learning reconstructions \label{sec:GA}}
We also consider a machine learning approach that can be used for non-parametric reconstruction of a given data set, called the Genetic Algorithms (GA). The GA follow a stochastic approach based on the genetic operations of crossover and mutation, in order to express the notion of grammatical evolution of a population of test functions applied to data reconstruction. The GA in particular emulate the notion of evolution via natural selection; a given population changes and adapts to its environment under pressure from the stochastic operators of crossover, i.e. a random change in the chromosomes of an individual, and mutation, i.e. the merging of different individuals to form descendants, usually called offspring. Then, the probability that the members of the population will produce offspring, or equivalently their reproductive success, is assumed to be proportional to their fitness. The latter is a measure of how well the members of the population fit  the data and in our analysis we take this to be a standard $\chi^2$ statistic. For various applications to cosmology and more details on the GA see \cite{Akrami:2009hp,Bogdanos:2009ib,Nesseris:2012tt, Arjona:2019fwb,Arjona:2020kco,Arjona:2020doi,Nesseris:2010ep,Nesseris:2013bia,Sapone:2014nna}.

In a nutshell, the process to fit the LSST, DESI and ET data is the following. First, an initial group of functions, called the initial population, is created based on a set of orthogonal polynomials which are called the grammar. This step is crucial, as the choice of the grammar has been shown to directly affect the rate at which the GA code converges \cite{Bogdanos:2009ib}. This initial population is then set up in a manner such that both the duality parameter $\eta(z)$ and  $d_L(z)$ are encoded simultaneously by every member of the population. At this point we may also demand that the targeted functions that are to be reconstructed, i.e. $\eta(z)$ and $d_L(z)$, satisfy a set of initial conditions or physical priors. For example, these priors might be that the duality parameter satisfies $\eta(z=0)=1$ or that the luminosity distance today is zero, i.e. $d_L(z=0)=0$, but in our analysis we remain completely agnostic with respect to the expansion history of the Universe and we do not assume any specific model for it, as well as for the DDR deviation mechanism.

As mentioned earlier, the fitness of every member of the population is estimated with a usual $\chi^2$ statistic, using the LSST, DESI and ET  data. After that, the crossover and mutation stochastic operators are applied to a subset of the best-fitting functions, which are chosen via tournament selection \cite{Bogdanos:2009ib}. This procedure is subsequently repeated hundreds of times in order to ensure the convergence of the GA code. We also repeat the analysis with several different random seeds, so as to avoid  biasing the fit because of the choice of a specific random seed.

In order to provide error bounds on the reconstructions, we follow the approach of \cite{Nesseris:2012tt,Nesseris:2013bia}, where the error regions are estimated using a path integral calculation over the functional space scanned by the GA. This approach was compared against bootstrap Monte Carlo error estimates and its accuracy was thus validated \cite{Nesseris:2012tt}. The specific numerical implementation of the GA we use in our analysis is based on the publicly available code \texttt{Genetic Algorithms}\footnote{\url{https://github.com/snesseris/Genetic-Algorithms}}.

\subsection{Results \label{sec:results-GA}}
In the GA approach we reconstruct the quantities  $d_L^\textrm{bare}(z)$, $\eta_\textrm{EM}(z)$ and $\eta_\textrm{GW}(z)$, where the latter two are defined as
\ba
\eta_\textrm{EM}(z)&=& \frac{d_L^\textrm{EM}(z)}{d_L^\textrm{bare}(z)}, \\
\eta_\textrm{GW}(z)&=& \frac{d_L^\textrm{GW}(z)}{d_L^\textrm{bare}(z)},
\ea
and  $\eta_\textrm{GW}(z)$ can also be related to the effective Newton's constant $G_\textrm{eff}(z)$ via \autoref{eq:dLGeff} as
\be
\eta_\textrm{GW}(z)= \sqrt{\frac{G_\textrm{eff}(z)}{G_\textrm{eff}(0)}}.
\ee
Note that in GR and the \lcdm model, both $\eta_\textrm{EM}(z)$ and $\eta_\textrm{GW}(z)$ are exactly equal to unity, hence any deviation from that value would hint towards new physics, either in the EM or MG sector respectively.

\begin{figure*}
    \centering
    \includegraphics[width=0.49\textwidth]{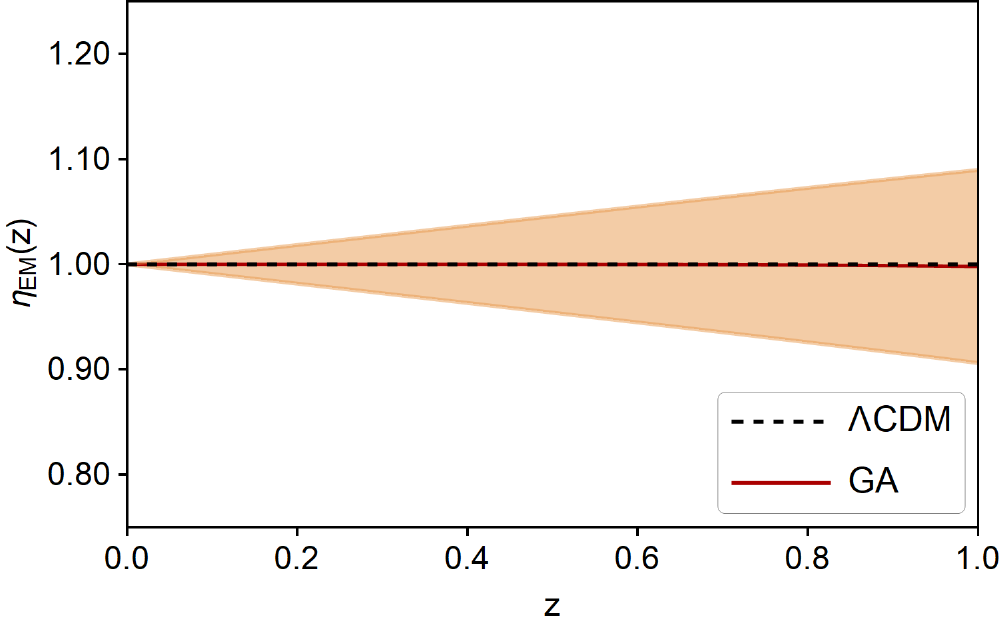}
    \includegraphics[width=0.49\textwidth]{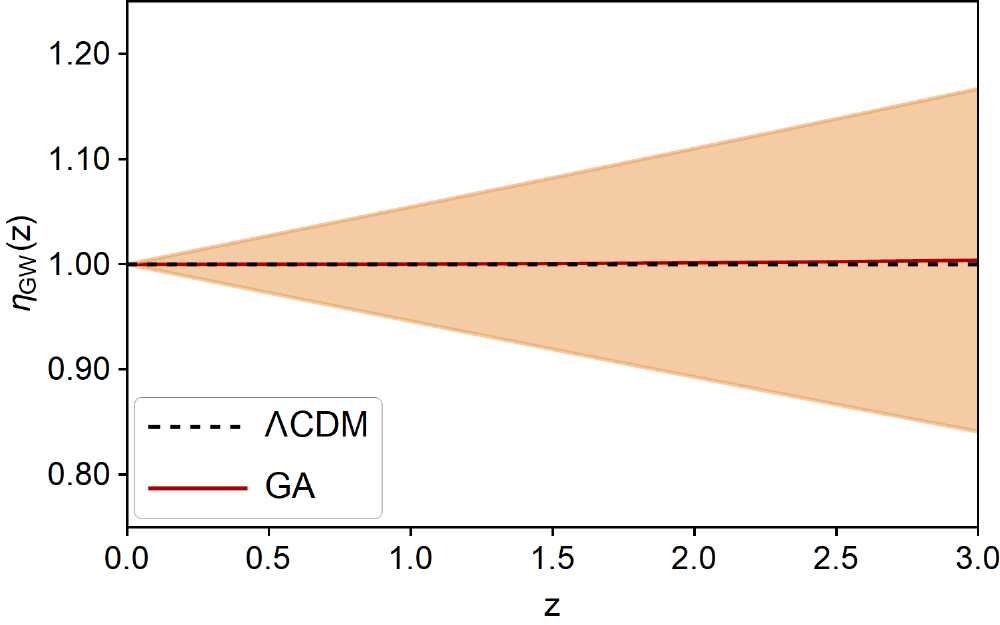}\\
    \caption{The GA reconstruction of $\eta_\textrm{EM}(z)$ (left) and $\eta_\textrm{GW}(z)$ (right) using the \lcdm mock. In both cases the orange shaded region corresponds to the $1\sigma$ GA  errors.\label{fig:GALCDMmocks}}
\end{figure*}

Having analysed the LSST, DESI and ET data with the GA, in what follows we now present the reconstructions of the two duality parameters. First, in \autoref{fig:GALCDMmocks} we show the GA reconstruction of $\eta_\textrm{EM}(z)$ (left) and $\eta_\textrm{GW}(z)$ (right) using the \lcdm mock, while the orange shaded region corresponds to the $1\sigma$ GA errors. As expected, in both cases the reconstructions are in perfect agreement with unity within the errors and the mean value of the GA follows exactly the fiducial value of the mock.

\begin{figure*}
    \centering
    \includegraphics[width=0.49\textwidth]{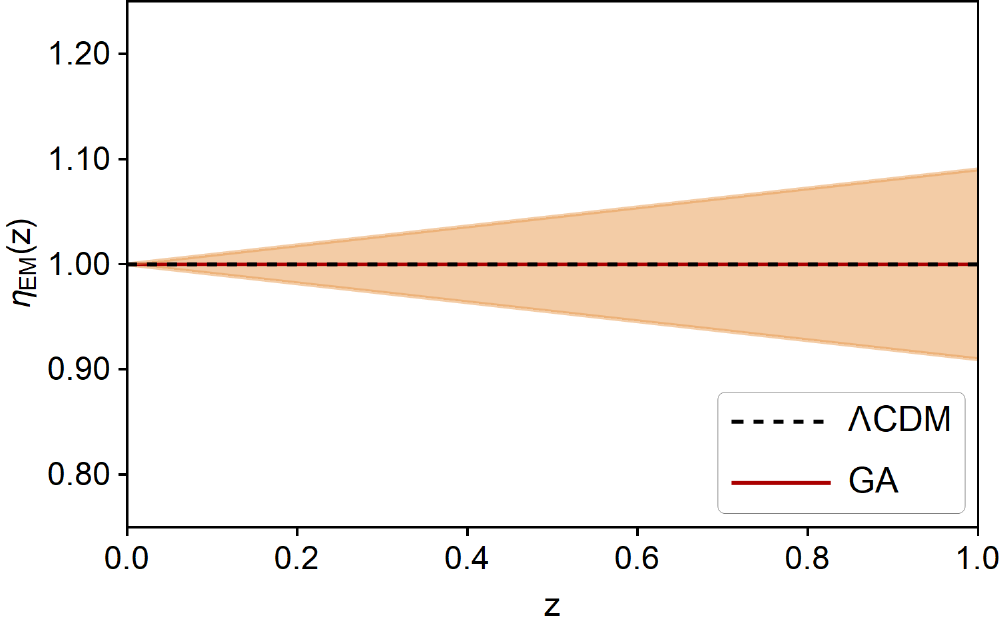}
    \includegraphics[width=0.49\textwidth]{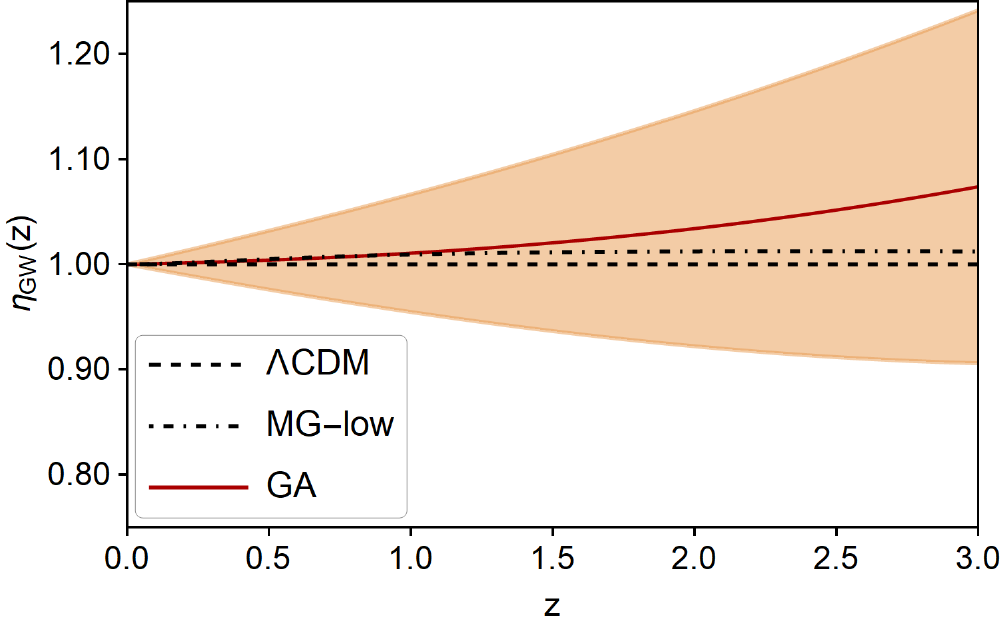}\\
    \caption{The GA reconstruction of $\eta_\textrm{EM}(z)$ (left) and $\eta_\textrm{GW}(z)$ (right) using the MG-low mock. The dot-dashed line corresponds to the fiducial model, while the orange shaded region corresponds to the $1\sigma$ GA errors. \label{fig:GAMGlowmocks}}
\end{figure*}

\begin{figure*}
    \centering
    \includegraphics[width=0.49\textwidth]{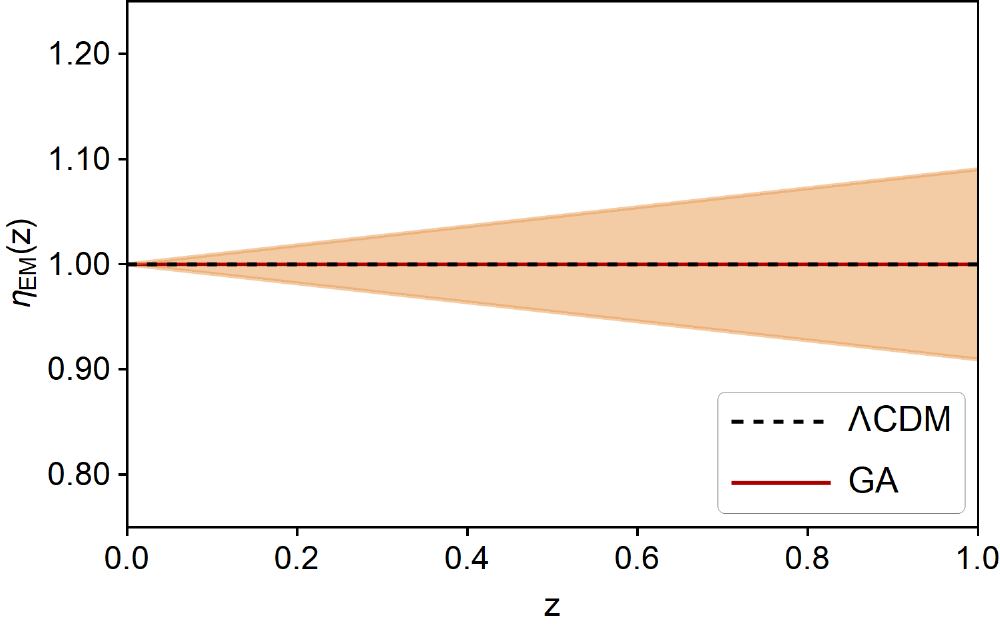}
    \includegraphics[width=0.49\textwidth]{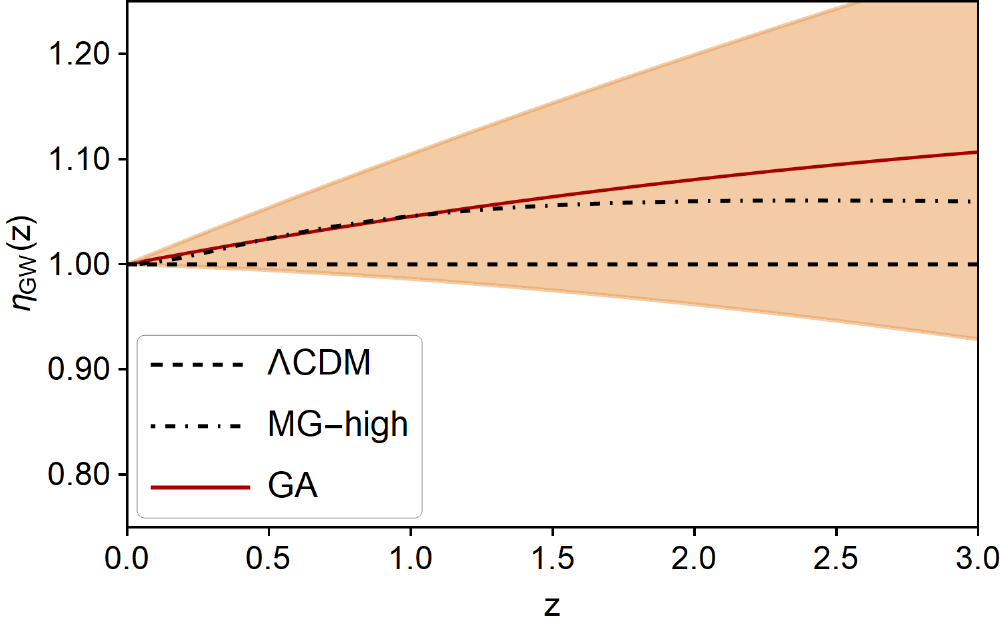}\\
    \caption{The GA reconstruction of $\eta_\textrm{EM}(z)$ (left) and $\eta_\textrm{GW}(z)$ (right) using the MG-high mock. The dot-dashed line corresponds to the fiducial model, while the orange shaded region corresponds to the $1\sigma$ GA errors.\label{fig:GAMGhighmocks}}
\end{figure*}

Next, we examine the results from the MG-low and MG-high mocks. In particular, we show the results of the GA reconstruction of $\eta_\textrm{EM}(z)$ (left) and $\eta_\textrm{GW}(z)$ (right) in \autoref{fig:GAMGlowmocks} and \autoref{fig:GAMGhighmocks} respectively for the two mocks. In these plots, we also show the theoretical value of the $\eta_\textrm{GW}(z)$ with a dot-dashed black line, using the values $n=2$ and $g_a=(0.1,0.5)$ for the MG-low and MG-high mocks respectively, while the orange shaded region corresponds to the $1\sigma$ GA errors. As can be seen, while the $\eta_\textrm{EM}(z)$ reconstructions are in agreement with unity, the mean value of $\eta_\textrm{GW}(z)$ follows perfectly the fiducial model until $z\sim1$ where the SNIa data end, albeit the parameter is consistent with unity within the errors in both cases. 

\begin{figure*}
    \centering
    \includegraphics[width=0.49\textwidth]{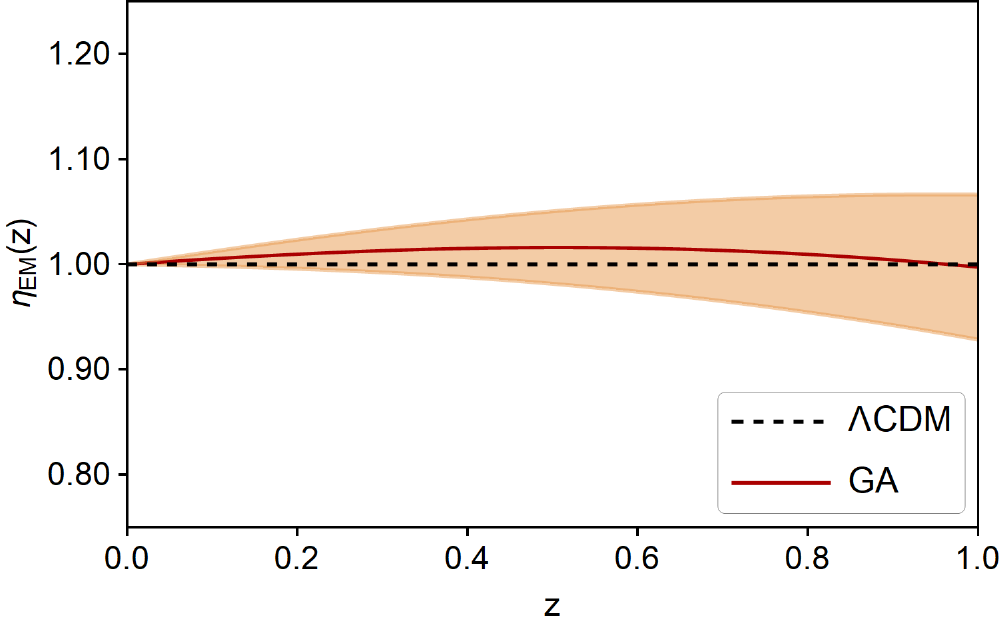}
    \includegraphics[width=0.49\textwidth]{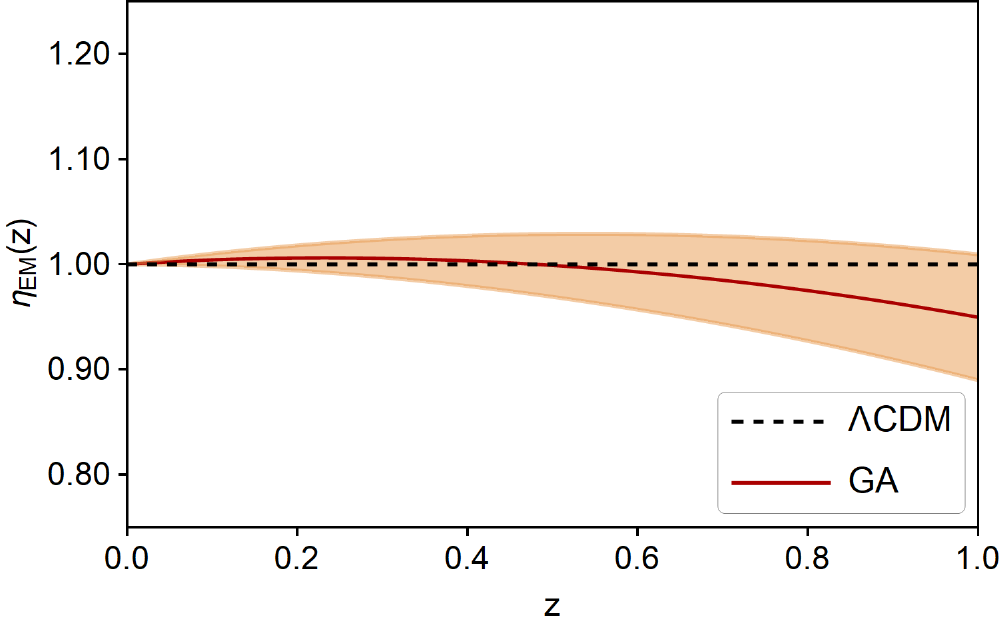}\\
    \caption{The GA reconstruction of $\eta_\textrm{EM}(z)$ using the LSST+DESI (left) and LSST+ET (right) combinations of the MG-high mock, while the orange shaded region corresponds to the $1\sigma$ GA errors. In both cases, we assume no MG effects in the likelihood.\label{fig:GAMGhighmocksmixed}}
\end{figure*}

Finally, in \autoref{fig:GAMGhighmocksmixed} we show the GA reconstruction of $\eta_\textrm{EM}(z)$ using the LSST+DESI (left) and LSST+ET (right) combinations of the MG-high mock, while the orange shaded region corresponds to the $1\sigma$ GA errors. In both cases, we assume that any possible deviation of the DDR is sourced from the EM sector, thus we assume that there are no MG effects, i.e. $G_\textrm{eff}=G_N$ and $\eta_\textrm{GW}(z)=1$, in order to examine any possible biases between the EM and MG sectors. As can be seen in the left panel of \autoref{fig:GAMGhighmocksmixed}, in the case of the LSST+DESI data combination, the GA reconstruction shows a deviation of the mean $\eta_\textrm{EM}(z)$ from unity as in the case of the SNIa alone any effect of the MG mocks can be reabsorbed in the luminosity distance, thus rescaling it and affecting the DDR relation. We also observe a similar behaviour for the other data combination, LSST+ET. Hence, in both cases we also confirm the finding of the parameterised approach that neglecting the MG effects in the likelihood leads to biases in the recovered quantities.

While some deviation from the fiducial value of $\eta(z)$ is found in these plots at high redshifts $(z\ge 1.5)$, and this is compatible with the results of the parameterised approach, there is no a priori widely accepted method on how to quantify it. In principle however, this could be done through a comparison of the $\chi^2$ values obtained for the GA and fiducial model functions.

\section{Conclusions\label{sec:conclusion}}

In this work we assessed the capability of future GW observations to constrain the DDR, alongside the commonly used observations of SNIa and BAO. Specifically, we investigated the constraining power of standard sirens both in combination with SNIa and BAO, and as an alternative to the latter, thus allowing us to overcome possible assumptions contained in the BAO data.

We firstly examined how the standard DDR can be broken by a mechanism in which photons decay into axions in the presence of magnetic fields, demonstrating how standard sirens can break degeneracies between parameters, and improve the constraints on cosmological parameters as well as constraining the axion model parameter $A$, which encodes the amplitude of the deviation from the standard DDR in this model.

We then explored how a generic modified gravity toy model with a time-varying Newton's constant can alter the gravitational luminosity distance measured by standard sirens, while at the same time affecting the measurement of the distance modulus through SNIa observations. Using two mock datasets with different strengths of modified gravity, MG-low and MG-high, we showed how the DDR breaking in the gravitational sector leads to a $4 \sigma$ false detection of DDR breaking in the electromagnetic regime via the photon--axion decay model in the case of the MG-low mock, if such modified gravity effects are not properly taken into account. The significance of the false detection rises to $10\sigma$ in the case of the MG-high mock. 

This false detection demonstrates the powerful effect on the results of the parameter estimation pipeline of the assumptions made when running an MCMC analysis. Consequently, we showed that the effect can be mitigated by including the modified gravity parameter $g_a$ in our analysis as a free parameter, although the strong degeneracy between this parameter, the axion model parameter $A$ and the matter density parameter $\Omega_{\rm m}$ meant that reasonable constraints were only obtained by the full combination of the data, LSST+ET+DESI.

It is well known that many modified gravity models require a screening mechanism in order to evade stringent solar system constraints. We investigated the consequences of screening on our constraints, considering a case in which only the GWs are screened, and a case in which only the SNIa are screened. In the first case, we found yet another false detection of DDR violation in the LSST+ET and LSST+DESI datasets, due to the presence of the unscreened SNIa in both. In the case of screened SNIa with unscreened GWs, we found that any combination which includes the ET mock data was biased away from the fiducial cosmology, revealing a potential smoking gun for the presence of modified gravity, if such an effect were to be observed in real data.

Finally, we performed a non-parametric reconstruction of the distance duality parameter $\eta(z)$ using a specific machine learning approach, based on the GA. We showed that the GA can correctly discriminate between the \lcdm, MG-low and MG-high mocks, as in the case of the latter two the mean GA value of the $\eta_\textrm{GW}(z)$ parameter shows deviations from unity and it follows the fiducial model perfectly until the range covered by the SNIa ($z\sim1)$. On the other hand, if we neglect the effects of modified gravity in the likelihood, then the reconstruction leads to biases as the GA cannot discriminate the MG from the EM effects, due to degeneracies in the parameters, something which is in agreement with the parameterised approach.

In conclusion, we have seen how mock datasets of standard siren events in combination with Type Ia supernovae and baryon acoustic oscillations are an excellent way to understand the potential constraining power of future surveys when applied to the distance duality relation. However, as we have shown, rigorous checks of all possible degeneracies and biases should be carried out when using standard siren data in combination with other probes, to ensure that no false detections of beyond $\Lambda$CDM physics are accidentally made. With this important finding in mind, it becomes abundantly clear that, as our GW detectors continue to improve and the number of observed BNS events begins to increase, standard sirens will become a vital part of all future cosmological analyses.

\acknowledgments
The authors would like to thank I.~Harry, K.~Koyama, C.J.A.P.~Martins, I.~Tutusaus and B.~S.~Wright for informative discussions and comments. Numerical computations were done on the Hydra HPC Cluster of the Instituto de F\'isica Te\'orica UAM/CSIC and the Sciama HPC cluster which is supported by the ICG, SEPNet and the University of Portsmouth. NBH is supported by UK STFC studentship ST/N504245/1 and gratefully acknowledges the hospitality of the IFT where part of this work was carried out. MM has received the support of a fellowship from ``la Caixa'' Foundation (ID 100010434), with fellowship code LCF/BQ/PI19/11690015, and the support of the Spanish Agencia Estatal de Investigacion through the grant “IFT Centro de Excelencia Severo Ochoa SEV-2016-0597”. MM also wants to thank the Big Star Bar for providing a work space and an internet connection during this period of remote work. SN acknowledges support from the research projects PGC2018-094773-B-C32, the Centro de Excelencia Severo Ochoa Program SEV-2016-059 and the Ram\'{o}n y Cajal program through Grant No. RYC-2014-15843. This paper is based upon work from COST action CA15117 (CANTATA), supported by COST (European Cooperation in Science and Technology).

\appendix

\section{Mock datasets\label{sec:mockdata}}

\subsection{Type Ia supernovae}\label{sec:snmock}
Here we present the details of the SNIa mocks used in our analysis. In particular, we simulate SNIa observations based on the specifications of the Legacy Survey of Space and Time (LSST), performed by the Vera C. Rubin Observatory  \cite{Abell:2009aa}. The LSST deep-drilling fields will observe $N_{\rm SNIa}=8800$ SNIa in the redshift range $z\in[0.1,1.0]$, for which we use the redshift distributions of \cite{Astier:2014swa}. Regarding the error budget of the observations, we follow \cite{Astier:2014swa} and for every event $i$ we assign an observational error $\sigma_{{\rm tot},i}$ given by
\begin{equation}
    \sigma_{\textrm{tot},i}^2=\delta \mu^2_i+\sigma^2_{\textrm{flux}}+\sigma^2_{\textrm{scat}}+\sigma^2_{\textrm{intr}}\,.
\end{equation}
We have assumed that the contributions to the error due to the flux, scatter, and intrinsic uncertainties described in the previous equation are given by ($\sigma_{\textrm{flux}} = 0.01$, $\sigma_{\textrm{scat}} = 0.025$, and $\sigma_{\textrm{intr}} = 0.12$) respectively and are the same for all events. However, we also include an error on the distance modulus $\mu=m-M$, \begin{equation}
    \delta\mu_i = e_M~z_i,
\end{equation}
which evolves linearly in redshift, but now the parameter $e_M$ is normally distributed with standard deviation $\sigma(e_M)=0.01$ and vanishing mean \cite{Gong:2009yk,Astier:2014swa}.

\begin{figure}
    \centering
    \includegraphics[width=0.7\textwidth]{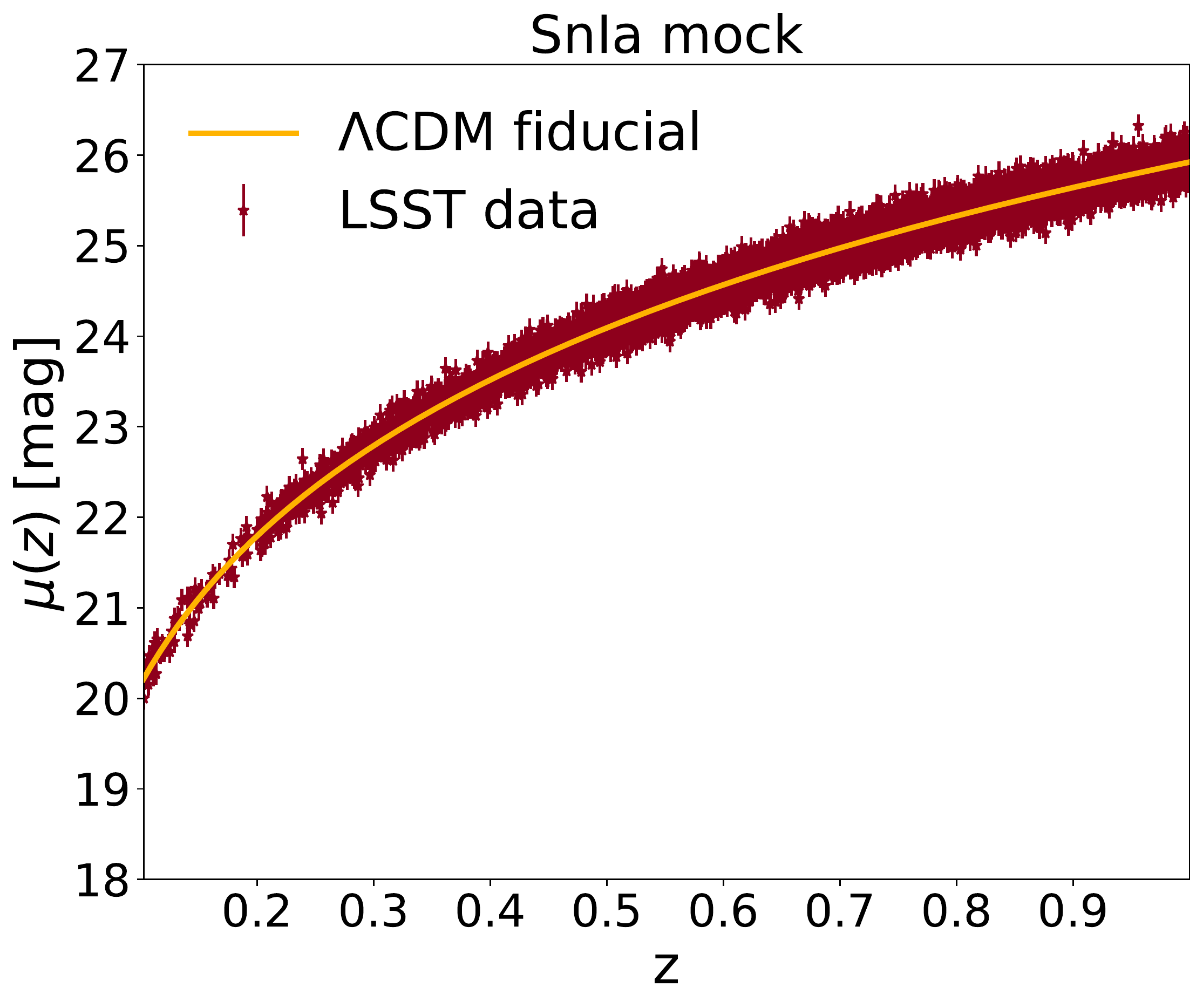}
    \caption{The distance modulus for the LSST $\Lambda$CDM SNIa  mock as a function of redshift. The data points are in red, while the fiducial is in yellow. The error-bars, correspond to $1\sigma$ errors.}
    \label{fig:mockSN}
\end{figure}

In \autoref{fig:mockSN} we show the distance modulus of the LSST $\Lambda$CDM SNIa mock, along with the $1\sigma$ error-bars of each point for the mock. The data points are in yellow, while the fiducial is in red. The error-bars, correspond to $1\sigma$ errors.

\subsection{Baryon acoustic oscillations\label{sec:baomock}}
For the baryon acoustic oscillation mocks, we make use of the extended redshift range of the Dark Energy Spectroscopic Instrument \cite{DESI2016}, which will probe the large scale structure and expansion rate of the Universe. The DESI survey will measure the optical spectra of tens of millions of quasars and galaxies up to $z\sim 4$, so as to enable redshift space distortion and BAO analyses. Here we base our mock DESI data on the official forecasts for both the angular diameter distance $d_{\rm A}(z)$ and the Hubble parameter $H(z)$ \cite{DESI2016}.\\ 
The DESI survey will have a coverage of approximately 14\,000\,deg$^2$ and the main types of DESI targets will be quasars, emission line galaxies (ELGs), luminous red galaxies (LRGs) and bright galaxies (BGs). The main DESI forecast measurements will cover the range $z\in [0.05,3.55]$, with a precision that may depend on the target population. In particular, the DESI BGs will be in the redshift range $z\in [0.05,0.45]$ in $5$ equispaced redshift bins, while the Ly-$\alpha$ forest quasar will be in the range $z\in [1.96,3.55]$ with $11$ equispaced redshift bins. On the other hand, the LRGs and ELGs will be in $z\in [0.65,1.85]$ with $13$ equispaced redshift bins. Finally, we also  assume that the aforementioned measurements will be uncorrelated.

\begin{figure}
    \centering
    \begin{tabular}{cc}
    \includegraphics[width=0.45\textwidth]{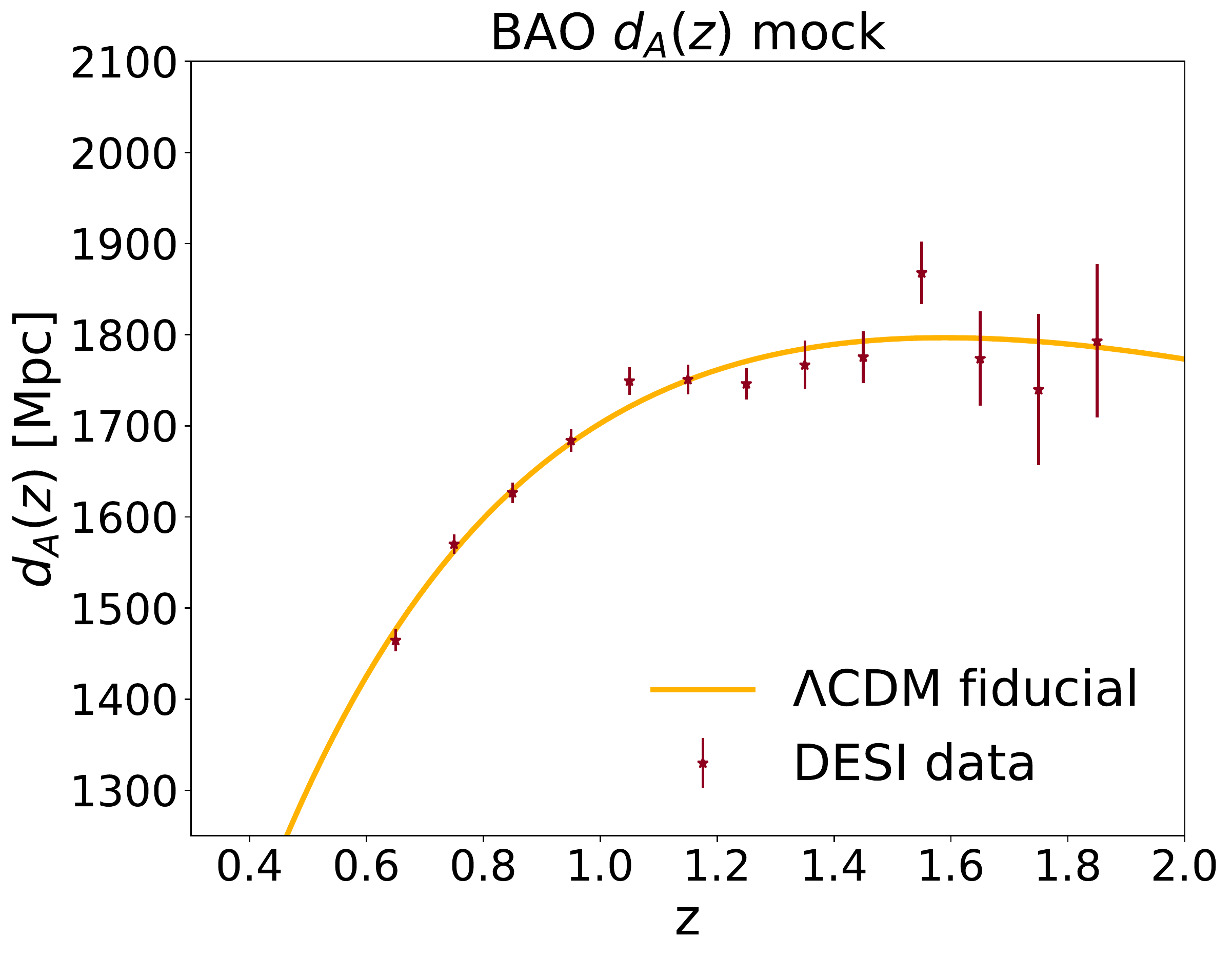} &
    \includegraphics[width=0.45\textwidth]{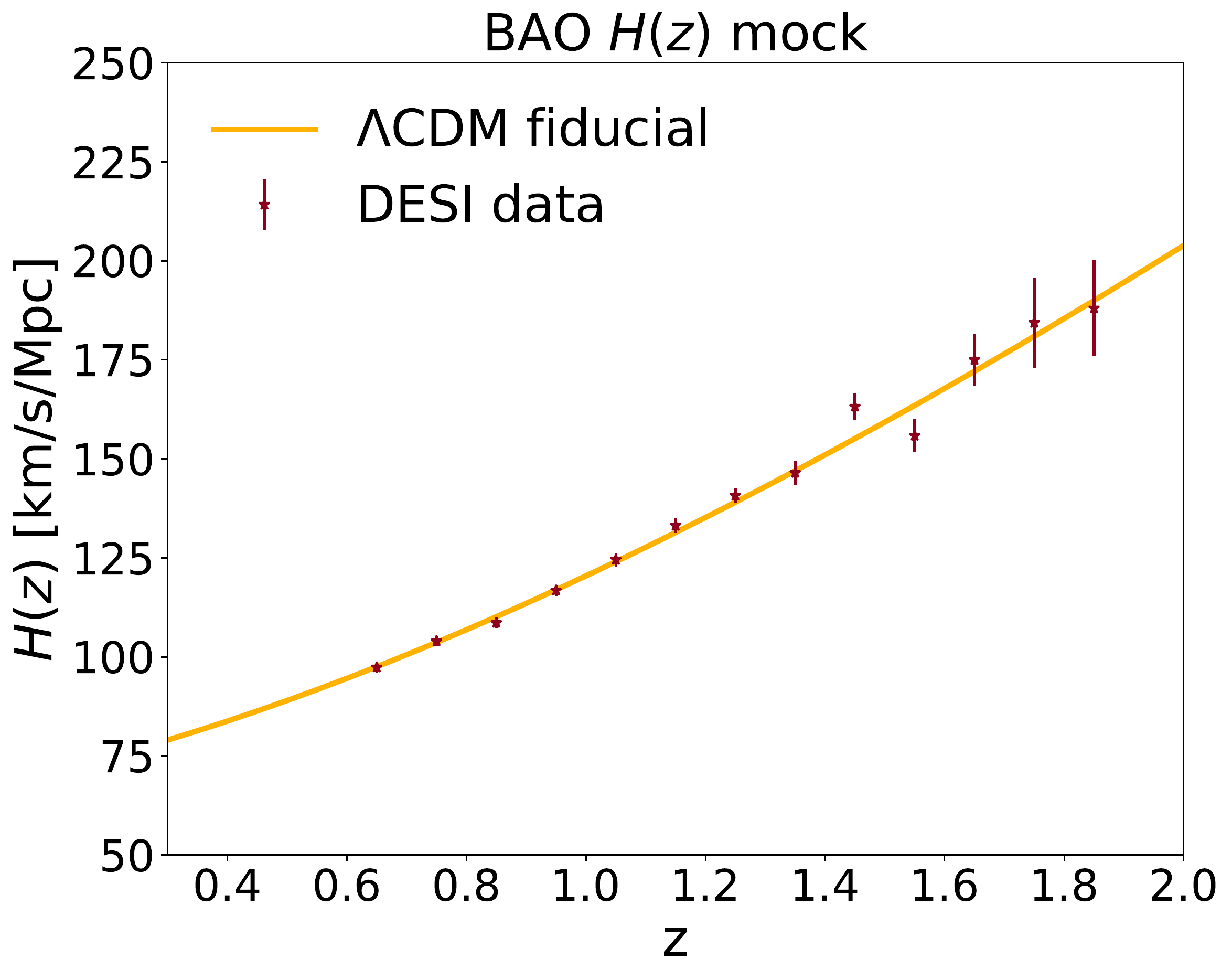}\\
    \end{tabular}
    \caption{The DESI $\Lambda$CDM BAO mocks for the angular diameter distance $d_A(z)$ (left) and the Hubble parameter $H(z)$ (right). The data points are in red, while the fiducial in each case is in yellow. The error bars, correspond to $1\sigma$ errors.}
    \label{fig:mockBAO}
\end{figure}

In our analysis in particular, we simulate measurements of the angular diameter distance $d_A(z)$ and the Hubble parameter $H(z)$ in the redshift range $z\in [0.05,3.55]$. In the left and right panels of \autoref{fig:mockBAO} we show the DESI $\Lambda$CDM BAO mocks for the angular diameter distance $d_A(z)$ (left) and the Hubble parameter $H(z)$ (right). The data points are in yellow, the fiducial model in each case is in red, while the error-bars, correspond to $1\sigma$ errors.

\subsection{Standard sirens}\label{sec:gwmock}
The inspirals and mergers of compact objects cause gravitational waves (GW) to propagate through spacetime. These waves can be detected by the strain $h(t)$ they produce in interferometers. This strain is expressed in the transverse traceless (TT) gauge as \cite{Maggiore2007}
\begin{equation}
    h(t)=F_+(\theta,\phi,\psi)h_+(t)+F_\times(\theta,\phi,\psi)h_\times(t),
\end{equation}
where $h_{+,\times}$ are the two independent components of the GW tensor $h_{\alpha\beta}$, $F_{+,\times}$ are the corresponding antenna pattern functions, $\psi$ is the polarisation angle, and $(\theta,\phi)$ is the angular position of the wave source on the sky with respect to the detector.

During an inspiral, there is negligible change in the orbital frequency over a single period. We can therefore compute the Fourier transform of the strain $h(t)$ in the stationary phase approximation \cite{Poisson1995,Sathyaprakash:2009xt, Zhao:2010sz},
\begin{equation}\label{eq:ftstrain}
    \mathcal{H}({\rm f}) = \mathcal{A}{\rm f}^{-\frac{7}{6}}\exp{\left[i(2\pi {\rm f}t_0-\frac{\pi}{4}+2\Psi({\rm f}/2)-\varphi_{(2,0)})\right]},
\end{equation}
where $t_0$ is a constant giving the fiducial epoch of the merger, which for the purposes of our analysis we set to zero. 

The Fourier amplitude $\mathcal{A}$ is  given by 
\begin{equation}\label{eq:amplitude}
    \mathcal{A}=\frac{1}{d_L^\textrm{GW}(z)}\sqrt{F_+^2[1+\cos^2(\omega)]^2+4F_\times^2\cos^2(\omega)}\sqrt{\frac{5\pi}{96}}\pi^{-\frac{7}{6}}\mathcal{M}_c^{\frac{5}{6}},
\end{equation}
where $\omega$ is the inclination of the orbital's angular momentum with respect to the line of sight, and $\mathcal{M}_c=M\eta^{3/5}$ is the chirp mass, related to the masses of the two binary component ($m_1$ and $m_2$) through the symmetric mass ratio $\eta=m_1m_2/M^2$ and the total mass $M=m_1+m_2$\footnote{Notice that the masses considered here are the observed masses, obtained from the intrinsic ones as $M_{\rm obs}=(1+z)M_{\rm int}$.}.

The phase $\Psi(\rm f)$ is given by
\begin{equation}\label{eq:phase}
    \Psi(\rm f) = \psi_0 + \frac{3}{256 \eta}\sum_{i=0}^{7} \psi_i (2 \pi M \rm f)^{(i/3)},
\end{equation}
where $\psi_0$ is the phase at the fiducial epoch and $\psi_i$ are the coefficients of the post-Newtonian expansion \cite{Blanchet:2013haa} (see Equation 129 of \cite{Sathyaprakash:2009xt} for the specific form used here). On the other hand, the function $\varphi_{(2,0)}$ is given by
\begin{equation} \label{eq:varphi}
\varphi_{(2,0)} = \tan^{-1} \left(-\frac{2 \cos(\omega) F_\times}{(1+\cos^2(\omega) F_+}\right).
\end{equation}

We can see from \autoref{eq:amplitude} that measuring the amplitude of GW signals allows estimates of the luminosity distances of the associated mergers to be obtained. However, in order to create a mock dataset of these mergers, we need to propagate the observational error to the luminosity distance. We focus here on the expected error for the Einstein Telescope, and following \cite{Cai:2016sby,Du:2018tia} we approximate the instrumental error on $d_L^{GW}$ as
\begin{equation}
    \sigma_{\rm inst}(d_L^{\rm GW})\approx \frac{2 d_L^{GW}}{\rho}, \label{eq:dlnoise}
\end{equation}
where $\rho$ is the combined signal to noise ratio of the three interferometers of the ET, with $\rho^2=\sum_i{\rho_i^2}$, and $\rho_i$ obtained as
\begin{equation}\label{eq:snr}
    \rho_i=\sqrt{\langle\mathcal{H}_i,\mathcal{H}_i\rangle}=\left[4\int_{{\rm f_{lower}}}^{{\rm f_{upper}}}{\mathcal{H}({\rm f})\mathcal{H}^*({\rm f})\frac{d{\rm f}}{S_h({\rm f})}}\right]^{\frac{1}{2}}.
\end{equation}
There is a correlation at play between the GW luminosity distance and the inclination of the source to the observer. For a single detector, $d_L^{\rm GW}$ and $\omega$ are completely degenerate with each other and the antenna patterns $F_{+,\times}$. However, with more than one detector, and sensitivity to both polarisations, this degeneracy can be broken. The maximum effect of this degeneracy on the signal to noise ratio is a factor of two, between the source being face on (inclination $\omega = 0$) and edge on ($\omega = \pi/2)$; this is the source of the factor of two that appears in \autoref{eq:dlnoise} \cite{Li:2013lza}.

In \autoref{eq:snr}, the function $S_h$ is the noise power spectral density; this function is provided for the ET in \cite{Zhao:2010sz}, where the antenna pattern functions for the three interferometers are given by 
\begin{eqnarray}\label{eq:antenna}
 F^{(1)}_+(\theta,\phi,\psi) &=& \frac{\sqrt{3}}{2}\left[\frac{1}{2}(1+\cos^2(\theta))\cos(2\phi)\cos(2\psi)-\cos(\theta)\sin(2\phi)\sin(2\psi)\right]\nonumber\\
  F^{(1)}_\times(\theta,\phi,\psi) &=& \frac{\sqrt{3}}{2}\left[\frac{1}{2}(1+\cos^2(\theta))\cos(2\phi)\cos(2\psi)+\cos(\theta)\sin(2\phi)\sin(2\psi)\right]\nonumber\\
  F^{(2)}_{+,\ \times}(\theta,\phi,\psi) &=& F^{(1)}_{+,\ \times}(\theta,\phi+\frac{2\pi}{3},\psi)\nonumber\\
  F^{(3)}_{+,\ \times}(\theta,\phi,\psi) &=& F^{(1)}_{+,\ \times}(\theta,\phi+\frac{4\pi}{3},\psi).
\end{eqnarray}

The frequency boundaries of \autoref{eq:snr} represent the cutoff frequencies of the observation. The upper one is connected to the last stable orbit (LSO) and is given by
\begin{equation}
    {\rm f}_{\rm upper}=2{\rm f}_{\rm LSO}=\frac{2}{6^{3/2}2\pi M_{obs}}.
\end{equation}
The lower cutoff frequency is dictated by the experimental configuration, as well as the location of the detector (for example, seismic noise affects ground-based detectors, raising this frequency in comparison to space-based detectors). We take the lower cutoff frequency to be $1$ Hz \cite{Sathyaprakash:2009xt,Zhao:2010sz}.

We now have all the equations needed to obtain the error $\sigma_{\rm inst}$ on the fiducial $d_L^{\rm GW}$ we computed. It is expected that the Einstein Telescope will be able to observe on the order of $10^5$ binary neutron star and neutron star-black hole mergers per year \cite{Sathyaprakash:2009xt}. However, only a small fraction of these will be accompanied by the visible optical counterpart necessary for cosmological parameter estimation. We therefore make the realistic assumption that 1000 binary neutron star (BNS) sources with an optical counterpart will be detected over a three year period. This allows us to make some other important simplifications:

\begin{itemize}
    \item We assume that these 1000 events are the subset of observations for which we can observe an electromagnetic counterpart in the form of short $\gamma$ ray bursts (SGRBs). This allows us to assume that we have a precise determination of the redshift for each event;
    \item As discussed in \cite{Du:2018tia,Jin:2020hmc}, the detection of SGRBs implies that the systems are oriented approximately face on, which allows us to assume that the inclination $\omega\approx0$. It is noted in \cite{Li:2013lza,Yang:2017bkv} that the maximum inclination angle is around $20\degree$, but if the Fisher matrix is averaged over the inclination and polarisation $\psi$ with the condition $\omega < 20 \degree$ this is the same as fixing the inclination to zero;
    \item We can assume the same masses of the binaries for all observed systems, with $m_1=m_2=1.4\ M_\odot$.
\end{itemize}

With these assumptions in mind we generate our $N$ events assuming a uniform distribution for their position $(\theta,\phi)$ in the sky, while for their distribution in redshift we use \cite{Zhao:2010sz}
\begin{equation}\label{eq:zdistr}
    P(z)\propto\frac{4\pi r^2(z)R(z)}{H(z)(1+z)},
\end{equation}
where $r(z)$ is the comoving distance and $R(z)$ is the merger rate of the binary systems, given by \cite{Cutler:2009qv}
\begin{equation}
    R(z) = \begin{cases} 1+2z &\text{if }z\leq1, \\ \frac{3}{4}(5-z) &\text{if }1<z<5, \\
    0 &\text{if } z\geq5.\end{cases}
\end{equation}

\begin{figure}
    \centering 
    \includegraphics[width=0.7\textwidth]{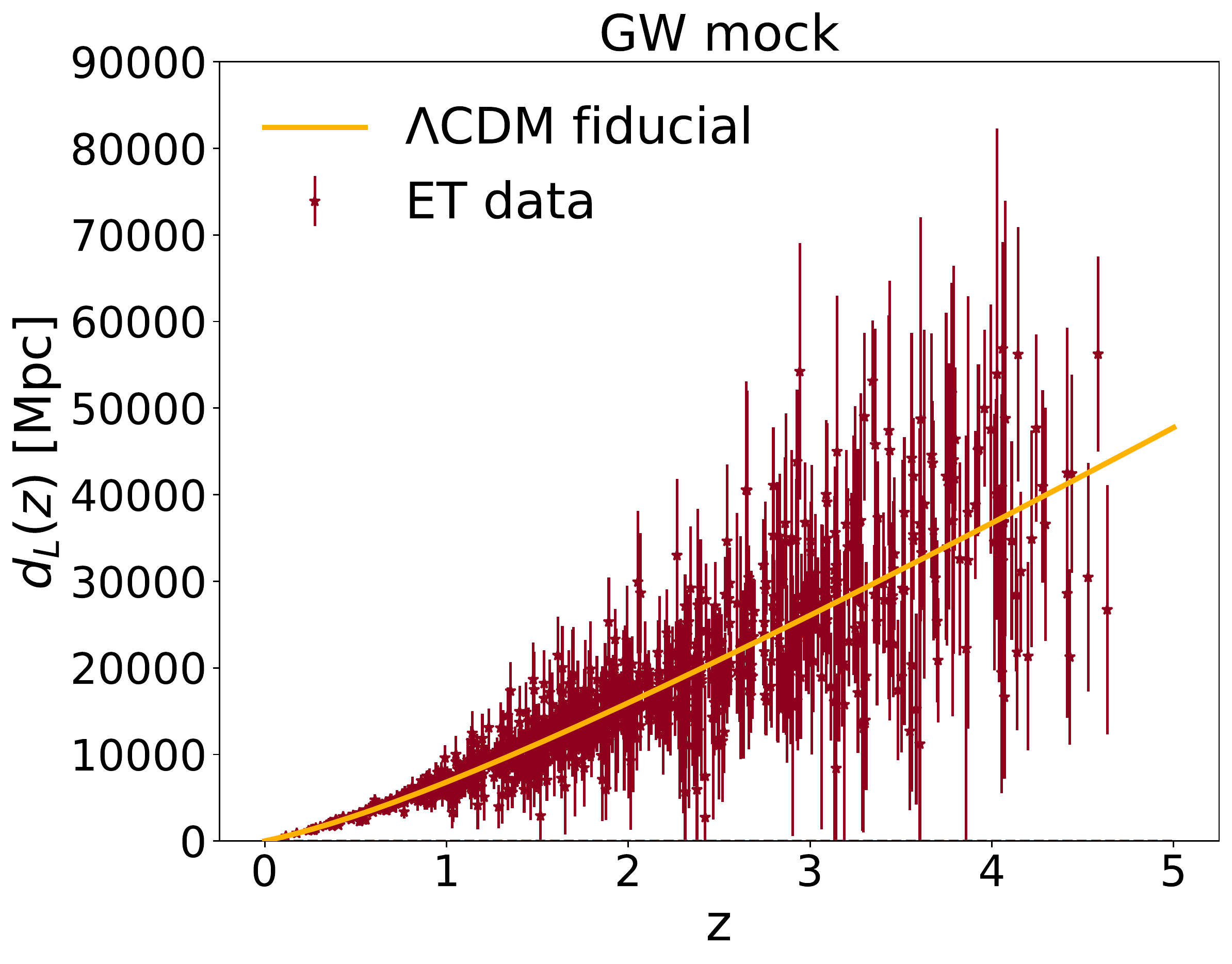} 
    \caption{Luminosity distance for the ET $\Lambda$CDM GW  mock as a function of redshift. The data points are in red, while the fiducial is in yellow. The error-bars correspond to $1\sigma$ errors.}
    \label{fig:mockGW}
\end{figure}

With the equations and the assumptions reported here, we are able to simulate our data points using the fiducial values of $d_L^{\rm GW}(z_i)$ at each sampled redshift $z_i$ and a total error on the observation given by
\begin{equation}
    \sigma(d_L^\textrm{GW})=\sqrt{\sigma_{\rm inst}^2+\sigma_{\rm lens}^2} \label{eq:error},
\end{equation}
where $\sigma_{\rm lens}\approx0.05~z~d_L^{\rm GW}$ is an extra error contribution given by weak lensing effects on the luminosity distance \cite{Sathyaprakash:2009xt}. Finally we simulate a spread of $d_L^{\rm GW}(z_i)$ with respect to the fiducial values, as it is given by observational noise, and therefore our final data points are obtained for each redshift from a Gaussian distribution with mean $d_L^{\rm GW}(z_i)$ and standard deviation $\sigma(d_L^{\rm GW}(z_i))$. The final result is shown in \autoref{fig:mockGW}.

\bibliographystyle{JCAP}
\bibliography{gwddr}

\end{document}